\newif\ifsubmode
\submodefalse

\newif\ifprintfig
\printfigtrue

\ifsubmode
  \documentstyle[12pt,aasms4,epsf,natbib209]{article}
  \received{}
  \accepted{}
  \journalid{}{}
  \articleid{}{}
\else
  \documentstyle[11pt,aaspp4,epsf,natbib209]{article}
\fi
 
\ifsubmode\else
\makeatletter
\def\thebibliography{\section*{References\@mkboth
 {REFERENCES}{REFERENCES}}\list
 {[\arabic{enumi}]}{\leftmargin 1em\labelwidth\z@\labelsep\z@\itemindent -1em
 \parsep -0.7ex
 \usecounter{enumi}}
 \def\newblock{\hskip .11em plus .33em minus -.07em}
 \sloppy\clubpenalty4000\widowpenalty4000
 \sfcode`\.=1000\relax}
\makeatother
\fi
 
\lefthead{Ferguson et al.}
\righthead{The Hubble Deep Fields}

\newcommand{\etal}{{et al.~}}

\newcommand{\gta}{\gtrsim}
\newcommand{\kms}{\:\rm km\,s^{-1}}
\def\lesssim{\mathrel{\hbox{\rlap{\hbox{\lower4pt\hbox{$\sim$}}}\hbox{$<$}}}}
\def\gtrsim{\mathrel{\hbox{\rlap{\hbox{\lower4pt\hbox{$\sim$}}}\hbox{$>$}}}}

\def\gtrsim{\mathrel{\hbox{\rlap{\hbox{\lower4pt\hbox{$\sim$}}}\hbox{$>$}}}}
\def\lesssim{\mathrel{\hbox{\rlap{\hbox{\lower4pt\hbox{$\sim$}}}\hbox{$<$}}}}
\def\U{U_{300}}
\def\B{B_{450}}
\def\V{V_{606}}
\def\I{I_{814}}

\def\Mpc{\:\rm{Mpc}}

\def\kpc{\:\rm{kpc}}

\def\msol{\:{M_{\odot}}}
\def\zphot{z_{\rm phot}}
\def\MER{\dot{\rho}_Z(z)}

\def\gta{\gtrsim}

\begin{document}

\title{The Hubble Deep Fields
\footnote{{\it Submitted for Annual Reviews of Astronomy and Astrophysics, vol. 38}}
}

\author{%
Henry C. Ferguson, 
Mark Dickinson and
Robert Williams \\
Space Telescope Science Institute, Baltimore, MD 21218\\
ferguson@stsci.edu
}


\ifsubmode

\noindent Send Proofs to: 

\begin{tabular}{l}
Henry C. Ferguson, Space Telescope Science Institute, \\
3700 San Martin Drive, Baltimore, MD 21218\\
Phone: (410) 338-5098 \\ 
Fax: (410) 338-4767\\
\end{tabular}
\fi


\ifsubmode
\clearpage
\fi

\begin{abstract} 
The Hubble space telescope observations of the northern Hubble deep field, and
more recently its counterpart in the south, provide detections and
photometry for stars and field galaxies to the faintest levels currently achievable,
reaching magnitudes $V \sim 30$. Since 1995, the northern Hubble deep field has been
the focus of deep surveys at nearly all wavelengths. These observations have
revealed many properties of high redshift galaxies, and have contributed
to important data on the stellar mass function in the Galactic halo.
\end{abstract}

\section{Introduction}

Deep surveys have a long history in optical astronomy. Indeed, nearly
every telescope ever built has at some point in its lifetime been
pushed to its practical limit for source detection. An important
historical motivation for such surveys has been the desire to test
cosmological models via the classical number--magnitude and
angular-diameter--magnitude relations \citep{Sandage61,Sandage88}.
However interpretation
has always been hampered by the difficulty of disentangling the effects
of galaxy selection and evolution from the effects of cosmic geometry.
The modern era of deep surveys in optical astronomy began with the
advent of CCD detectors, which allowed 4-m telescopes to reach to
nearly the confusion limit allowed by the point spread function
\citep{Tyson88}. These images revealed a high surface density of faint, 
blue galaxies, and the combination of their number--magnitude relation,
number--redshift relation, and angular correlation function was difficult
to reconcile with the critical-density universe $(\Omega_M = \Omega_{\rm tot} = 1)$ 
that seemed at the time to be a robust expectation from
inflationary models \citep{Kron80,FYTY90,GRV90,EBTKG91}.

Prior to launch, simulations based on fairly conservative assumptions suggested
that the Hubble space telescope (HST) would not provide an overwhelming advantage over ground-based
telescopes for studies of distant galaxies \citep{BGS90}:
\begin{quotation}
{\it Our results show that the most sensitive exposures achieved so far
from the ground reveal more galaxies per unit area than will be seen
by planned HST observations unless galaxy sizes decrease with the
maximum rate consistent with ground-based observations. In this,
the most favorable case for HST, the space exposures will show almost
as many galaxy images as have been observed so far in the most
sensitive ground-based data.}
\end{quotation}
It was apparent from the images returned immediately after the 
HST refurbishment in 1993 that distant galaxies had substantially
higher surface brightnesses than predicted by such simulations
\citep{DOSL94,Dickinson95Gal}.
Despite the pessimistic predictions, extensive field galaxy surveys
were always a top priority for HST, and large amounts of observing time
were awarded to deep, pointed surveys and to the medium-deep survey 
\citep{MDSIAU}. These
observations provided valuable information on galaxy scale lengths and
morphologies, on the evolution of clustering and the galaxy merger
rate, and on the presence of unusual types of galaxies \citep{MWSPGRCIEGGS94,SLCHLT95,CHS95Nat,MDSIAU}.
In this context, the idea of using the HST to do an ambitious deep field survey
began to look quite attractive, and was eventually adopted as one
of the primary uses of the HST director's discretionary time.

\subsection{This Review}

The Hubble deep fields north and south (HDF-N and HDF-S, respectively)
are by now undoubtedly among the most ``data rich''
portions of the celestial sphere.  The task of reviewing progress is
complicated by the fact that much of the supporting data are still being
gathered, and by the wide range of uses that have been found for
the HDF.  It is not practical to provide a comprehensive review of all HDF-related
research in a single review article. Instead
we have chosen to divide the review into two parts. The first part
discusses the data, both from HST and from other facilities, and
summarizes measurements and phenomenology of the sources found in
the field. The second part focuses specifically
on distant galaxies, and attempts to provide
a critical view of what the HDF has, and has not, taught us about
galaxy evolution. \citet{Ferguson98} has reviewed the HDF
with a different focus, and the series of
papers in the 1996 Herstmonceux conference and the 1997 STScI
May Symposium provide a broad summary of the overall field
\citep{Herst96,STScIMay97}. Throughout this review, unless 
explicitly stated otherwise, we adopt cosmological parameters
$\Omega_M, \Omega_\Lambda, \Omega_{\rm tot} = 0.3, 0.7, 1.0$ and 
$H_0 = 65 \kms {\rm Mpc^{-1}}$. Where catalog numbers of
galaxies are mentioned, they refer to those in \citet{WBDDFFGGHKLLMPPAH96}.

\section{Part 1: Observations, Measurements, and Phenomenology}

The HDFs were carefully selected
to be free of bright stars, radio sources, nearby galaxies, etc., and
to have low Galactic extinction.  The HDF-S selection criteria included
finding a quasi-stellar object (QSO) that would be suitable for studies of absorption lines 
along the line of sight.  Field selection was limited to the
``continuous viewing zone'' around $\delta = \pm 62^\circ$, because
these declinations allow HST to observe, at suitable orbit phases,
without interference by earth occultations. Apart from these
criteria, the HDFs are typical high-galactic-latitude fields;
the statistics of field galaxies or faint Galactic stars should be 
free from a priori biases.  The HDF-N observations
were taken in December 1995 and the HDF-S in October
1998. Both were reduced and released for study within 6 weeks of the
observations. Many groups followed suit and made data from
follow-up observations publicly available through the world wide web.

Details of the HST observations are set out in \citet{WBDDFFGGHKLLMPPAH96},
for HDF-N and in a series of papers for the
southern field \citep{WilliamsHDFS,FergusonHDFS,GardnerHDFS,CasertanoHDFS,FruchterHDFS,LucasHDFS}. The
HDF-N observations primarily used the WFPC2 camera, whereas the southern
observations also took parallel observations with the new instruments installed in
1997: the near-infrared camera and multi-object spectrograph (NICMOS)
and the space telescope imaging spectrograph (STIS).  The area of sky
covered by the observations is small: $5.3$ arcmin$^2$ in the
case of WFPC2 and $0.7$ arcmin$^{2}$ in the case of STIS and NICMOS for
HDF-S. The WFPC2 field subtends about 4.6 Mpc at $z
\sim 3$ (comoving, for 
$\Omega_M, \Omega_\Lambda, \Omega_{\rm tot} = 0.3, 0.7, 1.0$).
This angular size is small relative to scales relevant for
large-scale structure. 

The WFPC2 observing strategy was driven partly by the desire to
identify high-redshift galaxies via the Lyman-break technique
\citep{GTM90,SPH96}, and partly by considerations
involving scattered light within HST \citep{WBDDFFGGHKLLMPPAH96}. The images
were taken in four very broad bandpasses (F300W, F450W, F606W, and
F814W), spanning wavelengths from 2500 to 9000 {\AA}.  Although filter
bandpasses and zeropoints are well calibrated,\footnote{
None of the analyses to date have corrected for the WFPC2
charge-transfer inefficiency (CTE; \citealt{WHC99}). 
This is likely to be a small correction $\lesssim 5$ \% for the
F450W, F606W, and F814W bands. However, because of the low
background in the F300W band, the correction could be significantly
larger.}
no standard photometric
system has emerged for the HDF. In this review, we use the notation
$U_{300}, B_{450}, V_{606}$ and $I_{814}$ to denote magnitudes in the
HST passbands on the AB system \citep{Oke74}.  On this system $m(AB)
= -2.5 \log f_\nu({\rm nJy}) + 31.4$.  Where we drop the subscript,
magnitudes are typically on the Johnson-Cousins system, as defined
by \citet{Landolt73,Landolt83,Landolt92A,Landolt92B}
and as calibrated for WFPC2 by \citet{Holtzman2}; 
however we have not
attempted to homogenize the different color corrections and photometric
zeropoints adopted by different authors.

During the observations, the telescope pointing direction was shifted 
(``dithered'') frequently, so that the images fell on different detector
pixels. The final images were thus nearly completely 
free of detector blemishes, and were sampled at significantly higher
resolution than the original pixel sizes of the detectors. The 
technique of variable pixel linear reconstruction (``drizzling'') 
\citet{FH97} was developed for the HDF and is now
in widespread use. 

Both HDF campaigns included a series of shallow ``flanking field''
observations surrounding the central WFPC2 field. 
These have been used extensively to support
ground-based spectroscopic follow-up surveys. Since 1995 HDF-N has also
been the target of additional HST observations. 
Very deep NICMOS imaging and spectroscopy were carried out on
a small portion of the field by \citet{ThompsonHDF}.
\citet{DickinsonJdrop} took shallower NICMOS exposures to make a complete map of the
WFPC2 field.
A second epoch set of WFPC2 observations were obtained in 1997,
2 years after the initial HDF-N campaign \citep{GNP99}.
STIS ultraviolet (UV) imaging of the field is in progress, and a 
third epoch with WFPC2 is scheduled.

\ifsubmode\else

\begin{table}[h]
\begin{center}
\begin{tabular}{rl}
\multicolumn{2}{c}{\bf Table 1. Census of objects in the central HDF-N} \\
Number & Type of Source \\
\hline
$\sim 3000$ & Galaxies at $\U,\B,\V,\I$. \\
$\sim 1700$ & Galaxies at $J_{110}, H_{160}$. \\
$\sim 300$ & Galaxies at $K$ \\
9 & Galaxies at $3.2\mu$m \\
$\sim 50$ & Galaxies at 6.7 or 15$\mu$ \\
$\sim 5$ & Sources at $850\mu$m \\
$0$ & Sources at 450$\mu$m or 2800$\mu$m \\
6 & X-ray sources \nl
$\sim 16$ & Sources at 8.5 GHz \\
$\sim 150$ & Measured redshifts \\
$\sim 30$ & Galaxies with spectroscopic $z > 2$ \\
$<20$ & Main-sequence stars to $I = 26.3$ \\
$\sim 2$ & Supernovae \\
$0-1$ & Strong gravitational lenses \\
\hline
\end{tabular}
\end{center}
\end{table}

\fi

Table 1 gives a rough indication
of the different types of sources found in the HDF-N. 
Since 1995, the field has been imaged at wavelengths ranging
from $10^{-3}\,\mu$m (Chandra) to 
$2 \times 10^5\,\mu$m (MERLIN and VLA), 
with varying sensitivities, angular resolutions, and fields of view.
Not all of the data have been published or made public, but a large fraction 
have, and they form the basis for much of the work reviewed here.
Links to the growing database of 
observations for HDF-S and HDF-N are maintained on the 
HDF world wide web sites at STScI.


\section{Stars}

   The nature of the faint stellar component of the Galaxy is critical to
determining the stellar luminosity function and the composition of halo dark matter. 
In spite of the small angular sizes of the HDFs, their depth
enables detection of low-luminosity objects to large distances.
If the halo dark-matter is mostly composed of low-mass stars, there should 
be several in the HDF \citep{Kawaler96,Kawaler98p353}. 
Although HST resolution permits star-galaxy separation down to much
fainter levels than ground-based observations, many distant
galaxies also appear nearly point-like.  Color criteria are effective at
identifying likely red subdwarfs, but for $\I > 25$ and $\V-\I < 1$ the
possibility of galaxy contamination becomes significant. The counts of red
main-sequence stars have been compared with galactic models by
\citet{FGB96}, \citet{ESG96} and \citet{MMDMBC96}.
More detailed results have subsequently emerged from studies that have
combined the HDF-N with other HST images
to increase the sky coverage
\citep{GBF96,GBF97,RYMTS96,MG97,Kerins97,CM97}.  A
general conclusion is that hydrogen-burning stars with masses less than
0.3 $M_\odot$ account for less than 1\% of the total mass of the
Galactic halo. The overall HST database indicates that the Galactic
disk luminosity function experiences a decided
downturn for magnitudes fainter than $M_V = 12$  ($ M \lesssim 0.2 \msol$). 
For the halo, the constraints on the luminosity function are 
not as good because of the limited sky coverage. However, the 
luminosity function clearly cannot turn up by the amount it would need to
for main-sequence stars to be an important constituent of halo dark matter.

In addition to the nine or so candidate red dwarfs in the HDF-N, to a
limiting magnitude of $I = 28$ the field has about 50 unresolved objects
with relatively blue colors. Although in principle these could be young
hot white dwarfs (WDs), this possibility has appeared unlikely, as the
lack of brighter point sources with similar colors would require all
to be at very large distances, greater than 10 kpc away.
However, recent work by \citet{Hansen98,Hansen99} has shown that 
at low metallicity, molecular hydrogen 
opacity causes the oldest, lowest
luminosity WDs to become blue as they cool.  \citet{HDVHLSR99} have recently 
discovered such an object in the Luyten proper motion survey.
This result changes the way 
in which colors should be used to discriminate faint point-like sources in the
HDFs, and may lead to the reclassification of some faint galaxies as WDs.

Proper motions are an unambiguous way to distinguish between Galactic
stars and galaxies.  The HDFs serve as excellent first-epoch data for detection
of changes in position or brightness for any objects.  A second-epoch set of
images of the HDF-N were obtained 2 years after the
initial HDF-N campaign 
\citep{GNP99}, and
were analyzed by \citet{IRGS99} to search
for object motion in the images.
Of 40 identified point
sources with $I>28$, five were found to have proper motions that were $>3 \sigma$
above the measurement uncertainty ($\sim 10$ mas year$^{-1}$), with two of the 
objects having proper
motions exceeding 25 mas/yr and the remaining three near the detection limit.
The five objects are all faint ($I \sim 28$) and of neutral color ($V-I<0.9$), and
realistic velocities require that they have distances $d < 2$ kpc.  Although only
a very small fraction of the total sources in the HDF-N, these objects
represent a large increase over the number of stars expected from standard
models of the Galaxy, which predict less than one star in the range $27<V<29$
with $V-I<1.0$.  

\ifsubmode\else

\begin{figure}
\ifsubmode\else
\centerline{\plotfiddle{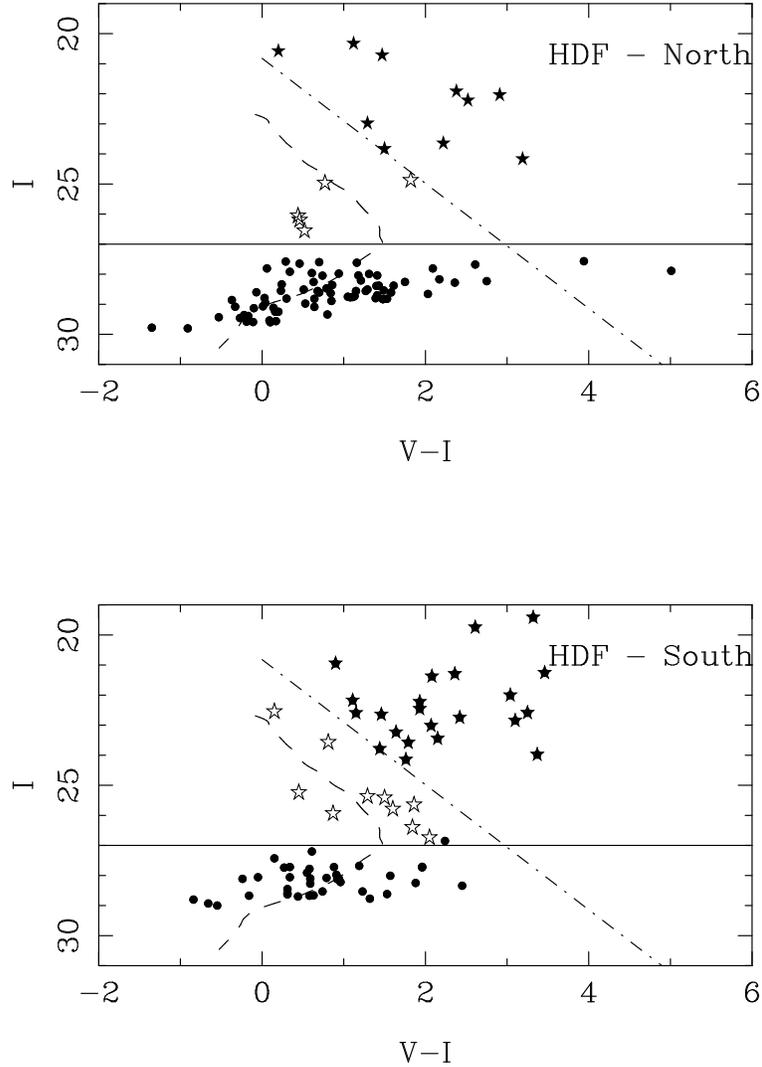}{6.0in}{0}{60}{60}{-400}{0}}
\fi
\caption{\label{fig_mendez}
Color-magnitude diagrams for point sources in the
Hubble deep fields (HDF).
({\it Dash-dot line}) Locus of disk M-dwarfs at a distance of 8 kpc;
({\it solid star symbols}) main-sequence disk and halo stars. ({\it Horizontal line})The ($15\sigma$) detection limit of $I=27$ magnitude; ({\it filled circles})
galactic stars and unresolved distant galaxies.
({\it Open star symbols}) Point sources which are too faint to be on the
main sequence at any reasonable distance, and which may be white dwarfs (WD).
({\it Dashed line}) The
$0.6 \msol$ WD cooling track from Hansen (1999) for a distance of 2 kpc.
(Adapted from Mendez \& Minniti 2000)
}
\end{figure}

\fi

Separate evidence may be emerging that is consistent with the
identification of faint blue stars in the HDFs.  The HDF-S ($l=328,
b=-49$) points closer to the Galactic center than HDF-N ($l=126,
b=55$), and hence samples a larger path length through the Galactic
halo. There should be more stars in HDF-S than in HDF-N.  \citet{MM99} found roughly double the number of blue point-like
sources in the HDF-S than its northern counterpart (Fig.
\ref{fig_mendez}).  This supports the hypothesis that a significant
fraction of these sources are stars.  A number of these stars are too
faint and too blue to be on the main sequence, and could
represent the old, low luminosity, blueish halo WDs
proposed by \cite{Hansen98}.

Although these developments are exciting, the evidence 
is far from compelling. The sources identified by \citet{MM99}
are brighter than those found to have proper motions by \citet{IRGS99}.
The two studies thus appear to be inconsistent, in that the closer WDs
ought to have larger proper motions. Also, the enhancement in faint
blue point-like objects in HDF-S relative to HDF-N is sensitive to the
magnitude limit chosen. If \citet{MM99} had included
objects down to $I = 29$ in their sample, they would have found more objects
in HDF-N than in HDF-S (Fig. \ref{fig_mendez}). 

Although still very tentative, the identification of WDs in the HDFs
is potentially extremely important. The results of the MACHO
project \citep{AAAABetal97} suggest that a substantial fraction of the
halo mass is due to objects with WD masses (although this is not a 
unique interpretation of the microlensing statistics) (\citealt{Sahu94}). 
If WDs do contribute significantly to the halo mass, then 
the early stellar population of the halo must have formed 
with a peculiar initial mass function (IMF) \citep{RYMTS96},
deficient in both high-mass stars [to avoid
over-enrichment of the halo by metals from supernovae (SN) and planetary
nebulae] and low mass stars, (because stars with $M < 0.8 \msol$ would still
be on the main sequence today). 
The presence or absence of WDs in deep fields will become more
definite within the next few years via forthcoming third-epoch
proper motion measurements.

\subsection{Supernova events}
 
An estimate of the SN rate at high redshift can provide important
constraints on the star-formation history of galaxies and on models
for SN progenitors \citep{PHDGGetal96,MDVP98,RC98}.
\citet{GP98} took second-epoch WFPC2 observations
of the HDF-N field in December 1997 and reported two SN
detections, at $I=26.0$ and 27.0.  \citet{MF99}
report a very faint detection, at $I=28.5$ of an object that brightened by
1 mag in 5 days in the original HDF-N images. The number of SNe
expected in a single epoch of HDF-N observations is 0.5-1
\citep{GNP99}, consistent with the number of detections.
The advanced camera for surveys (ACS) will provide substantial
benefits for future SN searches, because it will have roughly twice the solid angle
at five times the sensitivity of WFPC2. Based on the HDF numbers,
a survey of 12 ACS fields to HDF $I$-band depth should identify roughly 
25 SNe, and would cost $\sim 100$ orbits.

\section{Galaxies}


\ifsubmode\else
\begin{figure}
\ifsubmode\else
\centerline{\plotone{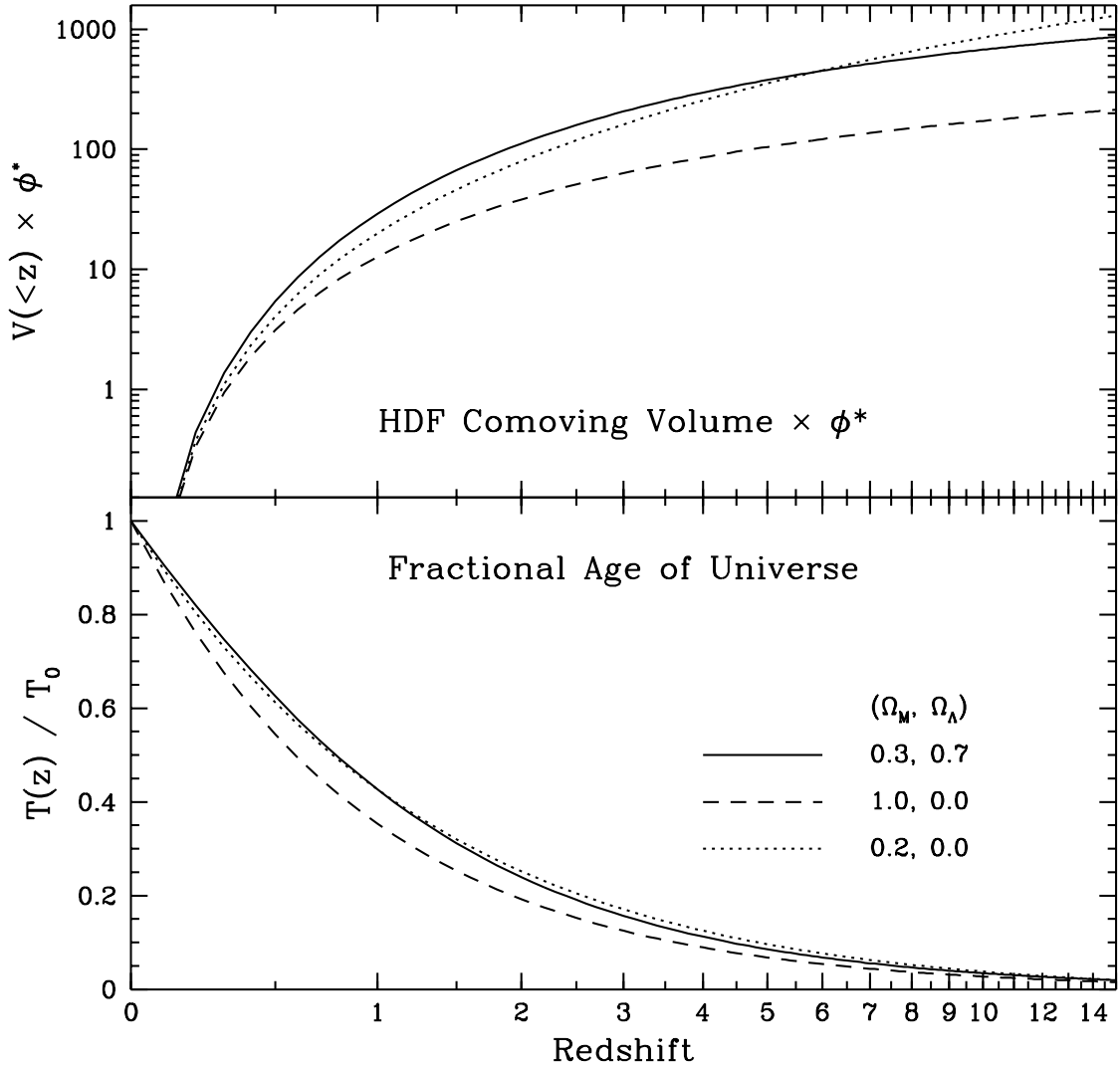}}
\fi
\caption{\label{fig_volume}
An illustration of volume and time in the Hubble deep fields (HDF).
({\it Top})
For the 5~arcmin$^2$ WFPC2 field of view, the co-moving volume out
to redshift $z$ is plotted for several cosmologies, scaled
by the present day normalization of the galaxy luminosity function
[$\phi^\ast$, here taken to be 0.0166$h^3$~Mpc$^{-3}$ from
Gardner (1997)].  This gives a rough measure of the
number of ``$L^\ast$-volumes'' out to that redshift.  ({\it Bottom})
The fractional age of the universe versus redshift is shown.
Most of cosmic time passes at low redshifts, where the HDF volume
is very small.}
\end{figure}

\nocite{GSFC97}
\fi

The HDFs are exquisitely deep and
sharp images, detecting thousands of objects distributed throughout
the observable universe, but they are also {\it very small} fields of
view, and each is only one sightline.  Figure \ref{fig_volume} illustrates 
some parameters relevant for studying galaxy evolution with 
a single HDF.  The total co-moving volume
out to redshift $z$ has been scaled ({\it top panel}) by the present-day
normalization of the galaxy luminosity function $\phi^\ast$.
This gives a rough measure of the number of ``$L^\ast$ volumes''
out to $z$, i.e.\ approximately the number of $L^\ast$ galaxies
expected in that volume, or at high redshift the number of
$L^\ast$ galaxies that the ``proto-galaxies'' found there 
will someday become.  At $z < 1$, this number is $\sim 10-30$
depending on the cosmology:  very few high-luminosity galaxies 
are expected (or found, for that matter), and even purely 
Poissonian variations introduce large uncertainties in any 
statistical conclusions that can be derived from them.
Given real galaxy clustering, these uncertainties are still
greater.  As an example, $\sim$24\% of the total rest-frame
6500\AA\ luminosity summed over all HDF-N galaxies out to 
$z = 1.1$ comes from just four galaxies:  two in a redshift 
``spike'' at $z = 0.96$ and two in another spike at $z = 1.02$.
The safest use (statistically) for the HDF at $z < 1$ is thus 
to study the vastly more numerous, low-luminosity galaxies.
This caution similarly applies to clustering studies. The angular
correlation functions derived from the $z \lesssim 1$ sample 
primarily refer to low-luminosity galaxies, whereas those at higher 
$z$ refer to higher luminosities. Clustering variation with
luminosity (or mass) can mimic evolution.

At high redshift, the HDF volume is large
(especially for the open and $\Lambda$ cosmologies), and thus is more 
likely to provide a fair sample of objects.\footnote{Even at high redshift, 
however, uncertainties due to clustering (cf.\ \citealt{ASGDPK98}) 
need to be kept in mind when analyzing one or two sightlines like the HDFs.}
For $(\Omega_M, \Omega_\Lambda) = (0.3, 0.7)$ there is $\sim 20\times$ more 
volume at $2 < z < 10$ than at $0 < z < 1$, and the likelihood of finding 
even moderately rare objects becomes significant.  There is, however, 
little {\it time} out at high redshift:  Most of cosmic history takes place 
at $z < 1$ (Figure 2, {\it bottom}).

\subsection{Galaxy colors and morphology}\label{sec_morph1}

One of the long-term goals of faint galaxy surveys is to chart the 
origin and history of the Hubble sequence. In the nearby universe, 
most galaxies near the knee ($L^*$) of the \citet{Schechter76} luminosity
function are elliptical, lenticular, or spiral galaxies. Although 
such normal galaxies dominate the integrated mass and luminosity density 
at $z = 0$, lower-luminosity dwarf irregular and dwarf elliptical galaxies
dominate by number. High-luminosity peculiar or interacting galaxies
are relatively rare, constituting 2-7 \% of the population in the local
volume $v < 500 \kms$ \citep{KM97}.

Attempts to extend galaxy classification and quantitative surface
photometry to magnitudes fainter than $V \sim 20$ were pioneered by
the medium deep survey (MDS) team \citep{CRGINOW95,DWOKGR95,GESG95}.
By the time of the HDF observations it was clear from such analysis
that the galaxy population at $I \sim 23$ contained a larger fraction
of irregular or peculiar galaxies than would be expected from a simple
extrapolation of galaxy populations in the local universe.  The HDF in
particular has spurred considerable effort toward quantifying galaxy
morphologies using a wide variety of methods, including traditional
Hubble-sequence classification, concentration and asymmetry
parameters, neural network clasification, and bulge/disk decomposition via
profile fitting \citep{vdB96,ATSEGvdB96,OWDK96,MS98}.  The classifications derived by different
methods are not always in agreement.  There are many considerations to
take into account when analyzing HST morphologies, including the behavior
of the classifier with signal-to-noise and image size (few HDF
galaxies have more than 100 independent resolution elements within an area
defined by some isophote reasonably exceeding the sky noise level), and
redshift effects both on surface brightness (the $(1+z)^4$ Tolman dimming)
and on rest-frame wavelengths sampled by a given image.    
\citet{ATSEGvdB96} used the HDF to extend the MDS results to $I=25$.
They measured the concentration and asymmetry of the galaxy images, and
also classified each galaxy by eye, finding that at $I=25$,
irregular/merging/peculiar systems comprise 40\% of all galaxies.  The
progressive shifting of UV wavelengths into the optical for higher
redshift galaxies may account for part of the trend \citep{BCHHLONSS91,GLBMS96,HV97}, but according to \citet{Abraham97p195} and
\citet{AEFTG99} this is not the dominant effect (see also 
\S\ref{sec_morph2} below).


\ifsubmode\else

\begin{figure}
\ifsubmode\else
\centerline{\plotone{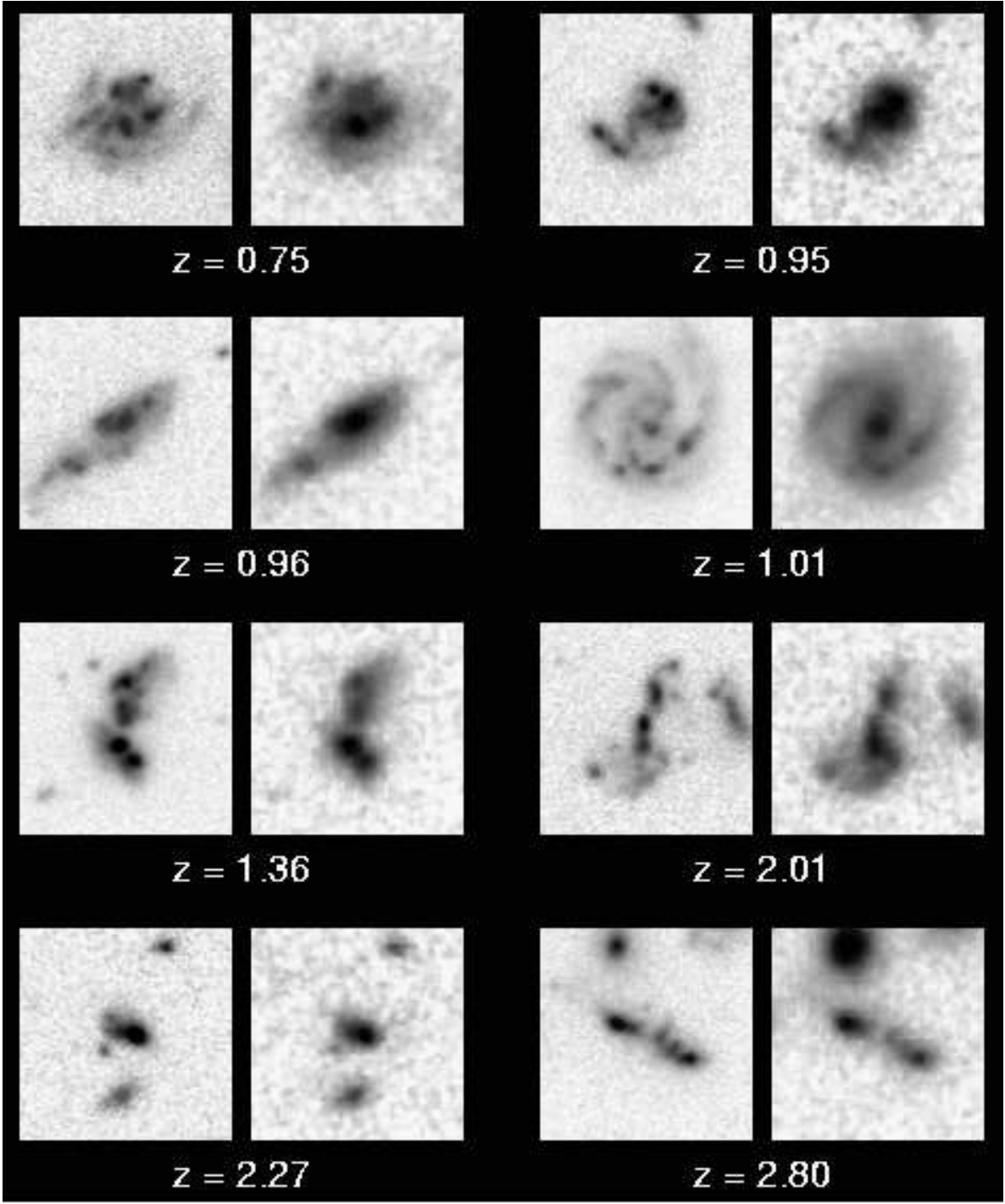}}
\fi
\caption{\label{fig_morphology}
Selected galaxies from the Hubble deep field north, viewed at optical and near-infrared
wavelengths with WFPC2 and NICMOS.  For each galaxy, the
color composites are made from $B_{450}$, $V_{606}$, $I_{814}$ ({\it left panel})
and $I_{814}$, $J_{110}$, $H_{160}$ ({\it right panel}).  The NICMOS images have
poorer angular resolution. In the giant spiral at $z = 1.01$, a prominent {\it red bulge}
and {\it bar} appear in the NICMOS images, whereas the spiral arms and HII regions
are enhanced in the WFPC2 data.  For most of the other objects, including
the more irregular galaxies at $z \sim 1$ and the Lyman Break galaxies
at $z > 2$, the morphologies are similar in WFPC2 and NICMOS.
}
\end{figure}

\fi

NICMOS observations (\citealt{TGMH98,ThompsonHDF,DickinsonHDFNIC,FruchterHDFS}), which sample the redshifted rest-frame optical
light out to nearly $z \sim 3$, further confirm that the observed changes 
in morphology with redshift are physical rather than an artifact of
bandpass shifting.  A few examples are shown in Fig. \ref{fig_morphology}
(see color insert).
Although interesting and occasionally dramatic
morphological transformations are found between the WFPC2 and NICMOS
images of some HDF galaxies, for
the large majority of objects with peculiar or irregular WFPC2
morphologies, the peculiarities persist in the NICMOS images.  The
structure of dramatic ``chain galaxies'' like 2-736.1 ($z = 1.355$) in
the HDF-N is almost entirely unchanging from 1300 {\AA} to 6800 {\AA} in
the rest frame.  \citet{CHS95} had noted that these
peculiar structures were unlikely to be stable or persist for long
given the nominal dynamical timescales for these galaxies, and
suggested that therefore they must be inherently young objects.  For
very blue galaxies like 2-736.1 this is likely to be true:  The
observed light from the UV through the IR is apparently dominated by
the same, relatively young generation of stars, and if there is an
older stellar component distributed differently then its visible light is
swamped by the younger stars. 

Although the more dramatic peculiarities in HDF galaxies generally persist
from optical through IR wavelengths, there do nevertheless appear to
be some general morphological trends with wavelength.   Bulges and bars are found
in NICMOS images of some $z \gtrsim 1$ disk galaxies that are invisible in the WPFC2
data, and disks appear to be smoother.  From a preliminary analysis
of structural parameters computed by \citet{CBDFPetal00}, we find that at
$H_{160} \lesssim 24$, galaxies tend to have smaller half-light radii, to be more
centrally concentrated, and to exhibit greater symmetry in the NICMOS images than in
their WFPC2 counterparts.  At fainter magnitudes, any trends are more difficult
to discern because the mean galaxy size in the NICMOS data becomes small enough
that the point-spread function (PSF) dominates structural measurements.  
\citet{TGMH98} noted similar
trends in the distributions of structural parameters from NICMOS parallel images.

Abraham et al. \citeyearpar{AEFTG99,Abraham_barred99} 
have studied a variety of other issues of galaxy morphology, 
concluding that the
fraction of barred spirals declines significantly for $z > 0.5$;
that $\sim 40\%$ of HDF-N elliptical galaxies at $0.4 < z < 1$
have a dispersion of internal colors that implies recent, spatially
localized star formation;  that spiral bulges are older than disks;
that the dust content of HDF spirals is similar to that locally; and
that the peak of past star formation for a typical $z \sim 0.5$ spiral
disk occurred at $z \sim 1$. These results are 
tentative because of small samples and uncertainties in stellar-population
models, but they show promise for future deep surveys.  

\subsection{Galaxy counts}

One of the benefits derived from the ready availability of the reduced
HDF data is that a variety of techniques have been used to catalog the
sources \citep{WBDDFFGGHKLLMPPAH96,LYF96,CROS96,SLY97,DAT96,Couch96,MSCFG96,MP00,CLP99,FontanaHDFS}.  Different algorithms have been used to construct the
catalogs, but all at some level rely on smoothing the image and
searching for objects above a surface brightness threshold set by the
background noise.  \citet{fergHDFsymp} compared a few of the
available catalogs and found reasonable agreement among them for 
sources brighter than $\V =
28$, with systematic differences in magnitude scales of less than 0.3
(nevertheless, there {\it are} systematic differences at this
level). The different catalogs apply different algorithms for splitting
and merging objects with overlapping isophotes. These differences,
together with the different schemes for assigning magnitudes to
galaxies, result in overall differences in galaxy counts. At $\I = 26$
the galaxy counts in the catalogs considered by
\citet{fergHDFsymp} all agree to within 25\%, whereas at $\I = 28$
there is a factor of 1.7 difference between them.  This highlights the
fact that galaxy counting is not a precise science.  


\ifsubmode\else
\begin{figure}
\ifsubmode\else
\centerline{\plotfiddle{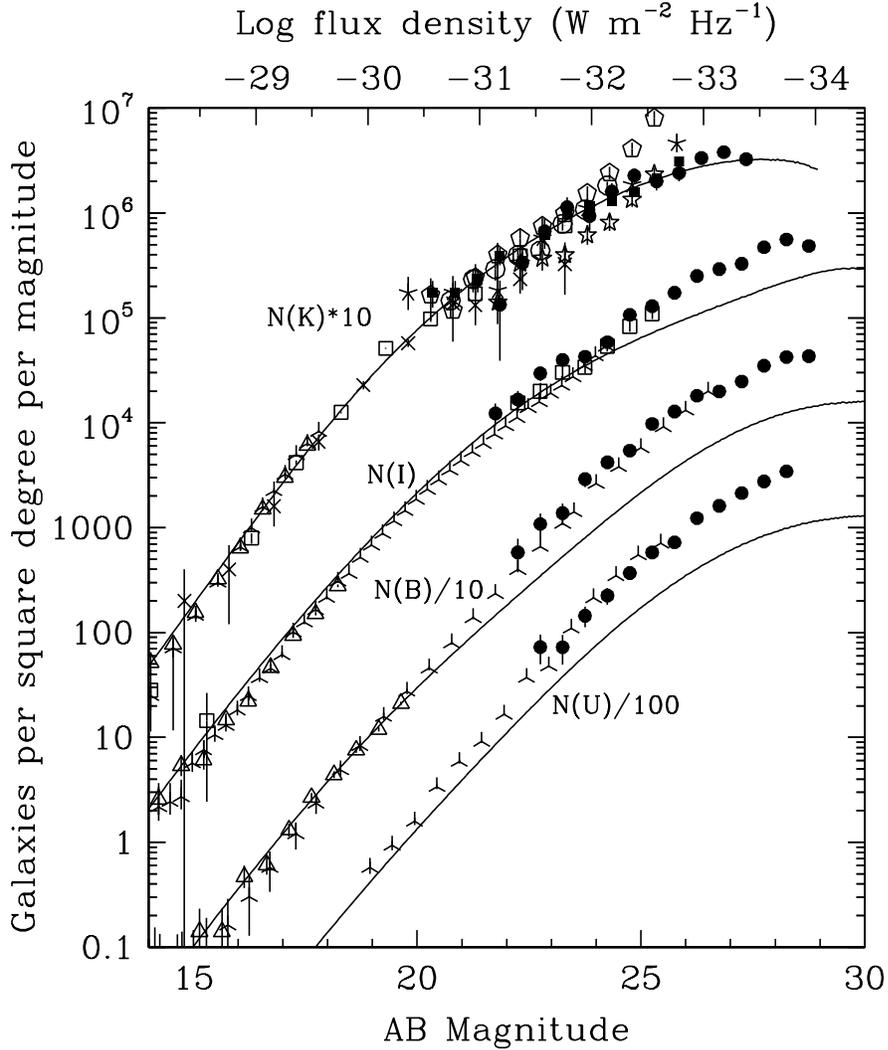}{5.8in}{0}{60}{60}{-400}{-20}}
\fi
\caption{\label{fig_counts}
Galaxy counts from the Hubble deep fields (HDF) and other surveys.
The HDF galaxy counts ({\it solid symbols}) use isophotal magnitudes and
have not been corrected for incompleteness. These corrections
will tend to steepen the counts at the faint end, but in a model-dependent
way. For the $K$-band, a color correction of $-0.4$ mag has been
applied to the NICMOS F160W band magnitudes. The HDF-N counts from
Thomson et al (1999) ({\it filled squares}) and the HDF-S
counts from Fruchter et al. 2000 ({\it filled circles}) are filled circles
are shown. For the
$U,B$ and $I$ bands, the HDF counts are the average of HDF-N
(Williams et al. 1996) and HDF-S (Casertano et al. 2000) with
no color corrections. The groundbased counts ({\it open symbols})
are from Mcleod \& Rieke (1995), Gardner et al. (1993; 1996),
Postman et al. (1998), Lilly Cowie \& Gardner (1991),
Huang et al. (1997), Minezaki et al. (1998), Bershady (1998), Moustakas et al. (1997),
and Djorgovski et al. (1995). ({\it Smooth curves}) A
no-evolution model with $\Omega_M,\Omega_\Lambda,\Omega_{\rm tot} =
0.3, 0.7, 0.1$ based on the luminosity functions, spectral-energy
distributions, and morphological type mix of the ``standard NE''
model of Ferguson \& McGaugh (1995).
}
\end{figure}

\nocite{MR95,LCG91,GCW93,GSCF96,HCGHSW97,MKYP98,BLK98,MDGSPY97,DSPLSNSMHBCHN95,FM95}

\fi

Figure \ref{fig_counts} shows the counts in four photometric bands
derived from the HDF and from ground-based observations.
As a fiducial comparison, a no-evolution model is shown for a spatially
flat model with $\Omega_M = 0.3$ and $\Omega_\Lambda = 0.7$. In all
bands the number--magnitude relation at HDF depths is significantly
flatter than $N(m) \propto m^{0.4}$, but at blue wavelengths it exceeds
the purely geometrical (no-evolution) predictions by roughly a factor of 3.
Interpretation of the counts is addressed in more detail in 
\S\ref{sec_counts2}.

\subsection{Galaxy sizes}

One of the interesting and 
somewhat unexpected findings of HST faint-galaxy surveys has been the
small angular diameters of faint galaxies. The sky is peppered
with compact high-surface-brightness objects, in contrast to
the expectation from $\Omega_M = \Omega_{tot} = 1$ pure-luminosity
evolution (PLE) models (\S\ref{sec_counts2}) 
and models dominated by low-surface-brightness
(LSB) galaxies \citep{FM95}.

Galaxy radii can be measured using image
moments (which are subject to severe biases because of isophotal
thresholds), growth curves, or profile fitting.  \citet{fergHDFsymp}
considered the size distribution derived from image moments, whereas
\citet{RRGIN98} discuss the distribution of half-light radii derived
from profile fitting. \citet{RRGIN98} find no evidence for evolution
in the rest-frame sizes or surface brightnesses of normal spirals and
ellipticals out to $z=0.35$.  At higher redshift they find evidence for
strong evolution; galaxies are more compact and of higher surface
brightness than expected from a PLE model (see also \citealt{SchadeCFRS95}).
The sizes of disk galaxies
were shown to be reasonably well matched by a size-luminosity evolution
(SLE) model wherein spirals form stars gradually from the inside out.
In a study with similar goals, \citet{SKFSVPGIW99} analyzed a sample of
190 (non-HDF) field galaxies with spectroscopic redshifts and HST
images. They find no evidence for size or luminosity evolution of
disk-dominated galaxies out to $z \sim 1$. The \citet{SKFSVPGIW99}
analysis includes a more comprehensive treatment of selection biases
and an empirical comparison to local samples that makes the results
quite compelling. However, \citet{SKFSVPGIW99} do identify nine 
high-surface-brightness objects in their sample at $z > 0.9$. If these
were included as disk galaxies, the results would be closer to 
those of \citet{RRGIN98}. A comparison
of the bivariate distribution of the \citet{SKFSVPGIW99}
luminosities and scale lengths 
to a $z \sim 0$ sample suggests only moderate density evolution
out to $z \sim 1.2$, with a decrease in the number density of
bright, large-scale-size galaxies at higher redshift
\citep{dJL99}. Galaxies in the
HDF with spectroscopic or photometric redshifts greater than $z = 2$
generally appear to be more compact than present-day $L^*$ galaxies
\citep{LickHDFhighz,RRGIN98}. 

At magnitudes fainter than $I = 25$, morphological classification and
profile fitting become quite difficult because most galaxies 
at these faint magnitudes are so small that even with HST 
there are few independent resolution elements within the area detected
above the background. \citet{fergHDFsymp} studied
the distribution of first-moment radii and modeled the selection
boundaries of the HDF in size and magnitude. 
Because the first-moment radii are measured above a fixed
isophote, the interpretation of the observed trends is
highly model dependent, and is sensitive to the assumed
redshift distribution and morphological-type distribibution
of the galaxies.
Nevertheless, down to $I \sim 27$
the HDF counts are clearly dominated by galaxies more compact than
local $L^*$ spirals. At fainter magnitudes, 
the locus of HDF galaxy sizes is tightly
constrained by selection, and there is very little information on the
{\it intrinsic} size distribution of the objects.
 
\subsection{Spectroscopic and Photometric Redshifts}\label{sec_spectroscopy}

The HDF-N and flanking fields have been the focus of some of the most 
intensive and complete
spectroscopic redshift survey work on faint galaxies.  
Essentially all spectroscopic redshifts
in the HDF have come from the W.M. Keck Observatory and LRIS
spectrograph \citep{CCHSBHS96,CHBCHSS00,LickHDFhighz,LickHDFcompact1,SGDA96,SAGDP99,Dickinson98p219,ASGDPK98,SSBDLYPF98,WSBSCTS98,WWCPSS99,ZMD97,HCM98}. The most recent compilation 
is 92\% complete for $R < 24$ in the central HDF-N, and also 92\% complete
for $R < 23$ in the flanking fields \citep{CHBCHSS00}.
The median redshift is $z \approx 1$
at $R = 24.$ Altogether, including unpublished
redshifts from the Steidel group, more than 700 spectroscopic redshifts
have been measured in the immediate vicinity of the HDF-N,  with $\sim
150$ in the central HDF (WF and PC) alone.  This latter corresponds to
30 redshifts per arcmin$^2$, spanning $0.089 < z < 5.60$ (and a few
stars), a density unmatched by any other field galaxy survey.
At the same time, it is sobering to realize
that less than 5\% of the galaxies from the \citet{WBDDFFGGHKLLMPPAH96}
HDF-N catalog have spectroscopic redshifts.  The vast majority are
fainter than the spectroscopic limit achievable with present-day
telescopes.
For the HDF-S, the extant spectroscopy is more limited, consisting of 
about 250 redshifts in an irregular area surrounding the three HST
fields \citep{Getal00,Detal00,TDPCW99}.  
Three galaxies at $2.8 < z < 3.5$ have been confirmed by the VLT
(\citealt{Cristiani99}). Eleven HDF-S ISO sources have redshift
measurements from VLT IR spectroscopy \citep{RigopoulouVLT}.

The estimation of galaxy redshifts using broad-band colors
received a tremendous boost from the HDF 
observations and followup studies.  Indeed, the advent of the HDF marked 
a transition for photometric redshifts, which evolved from an experimental
exercise to a widely used technique.  
This came about for several reasons.   Perhaps foremost, as noted above, 
the WFPC2 HDF data readily detected galaxies as much as 100 times fainter than 
the practical spectroscopic limits of 10m telescopes.  Therefore, in order 
to interpret the properties of the vast majority of HDF galaxies, 
non-spectroscopic techniques for estimating redshifts are necessary.   
Fortunately, the HDF-N WFPC2 data set offered very-high-quality, 
four-band photometry from 0.3 to 0.8$\mu$m for thousands of faint 
galaxies, and the extensive followup observations that came later extended 
this to other wavelengths and wider fields of view.  
Rapid dissemination of data and source catalogs from the HDF and followup 
observations made it relatively easy for any investigator to test their photometric 
redshift method.  Indeed, it should be noted that many photometric redshift 
studies to date have been tested and calibrated {\it solely} using the 
HDF-N. 

Several groups (\citealt{SGDA96,LickHDFhighz,CC96,MFDGSF96,MPD98}) used two-color selection keyed to the passage
of the Lyman limit and Lyman~$\alpha$ forest breaks through the F300W
and F450W passbands to identify galaxy candidates at $z > 2$. Such
color-selected objects are variously referred to as Lyman-break
galaxies, or UV dropouts. The \citet{MFDGSF96,MPD98} studies were
seminal in combining color-selected HDF samples at $z \sim 3$ and 4
with previous (spectroscopic, non-HDF) studies at $z < 2$ to derive a
measure of the global history of cosmic star formation, a goal
subsequently pursued by many other groups using various kinds of HDF
photometric redshifts (see \S\ref{sec_lumden}).

More general approaches to photometric redshifts have either fit
redshifted spectral templates to the photometric data (e.g.,
\citealt{GH96,MRGE96,LYF96,SLY97,FLY99,BBBSR99} for the HDF-N), or have used generalized
polynomial fits of redshift vs.\ multi-band fluxes to a spectroscopic
training set (e.g., \citealt{CSDSB97,WBT98}).  \citet{HawaiiActiveCatalog} and
\citet{CSDSB97} were the first to include IR HDF photometry
when deriving photometric redshifts.  In principle, this is especially
useful in the $1 < z < 2$ range where it is difficult to measure
redshifts from optical spectroscopy due to the absence of strong,
accessible emission or absorption features, and where the strong
spectral breaks (e.g. at 4000\AA\ and the 912\AA\ Lyman limit) fall
outside the optical passbands.   Many other groups have since included
IR photometry in their photometric redshift work, using the KPNO 4m
$JHK$ images from Dickinson et al. \citep{FLY99},
or more recently HST NICMOS data \citep{ThompsonPHOTZ,BSCCD99p19,YLCFPYP00}.

\citet{Hoggblind} and \citet{CHBCHSS00} have carried out blind tests
of the accuracy and reliability of HDF photometric redshifts, comparing
photometric predictions by the various groups with spectroscopic
redshifts unknown to those groups when the predictions were made.   In
general, all of the photometric techniques are quite successful at $z
\lesssim 1.4$, where the spectroscopic training sets are extensive,
with the best prediction schemes achieving $| z_{\rm phot} - z_{\rm
spec} | / (1+z_{\rm spec}) \lesssim 0.05$ for $\gtrsim 90$\% of the
galaxies.  At $z > 1.9$ the results are also generally quite good, with
10 to 15\% RMS in $\Delta z / (1+z)$ after excluding the worst
outliers.  
The intermediate redshift range, where photometric redshifts are particularly 
interesting, has not yet been tested due to the lack of spectroscopic calibrators.

Photometric redshifts have been used to identify galaxy candidates
in both HDFs at $z > 5$.  Two objects noted by \citet{LYF96}
and \citet{FLY99} have been subsequently confirmed
by spectroscopy, with $z = 5.34$ and 5.60 \citep{SSBDLYPF98,WSBSCTS98}.  With the addition of near--infrared data,
it becomes possible, in principle, to push to still higher redshifts.
Indeed, at $z > 6.5$, galaxies should have virtually no detectable
optical flux.  Candidates at these very large redshifts have been
identified by \citet{LYF98} and \citet{YLCFPYP00}.
One of these in the HDF-N was readily detected in NICMOS 1.6$\mu$m
images by \citet{DickinsonJdrop}, but is not significantly detected
at $J_{110}$ or at optical wavelengths.  Although this is a plausible
candidate for an object at $z \gta 12$, there are also other possible
interpretations.   It is curious that several of the proposed $z > 10$
candidates, including this ``$J$-dropout'' object in the HDF-N and
several of the HDF-S/NICMOS candidates proposed by \citet{YLCFPYP00},
are surprisingly bright ($\sim 1\mu$Jy) at $2.2\mu$m.   These would be
significantly more luminous than the brightest known Lyman break galaxies
at $2 < z < 5$.  It remains to be seen whether this
points to some remarkable epoch of bright objects at $z > 10$, or whether
instead the photometric redshifts for these objects have been overestimated.

\subsection{Galaxy Kinematics}

Spectroscopy of galaxies in the HDF-N has provided kinematical
information leading to constraints on mass and/or luminosity
evolution.  
\citet{LickHDFcompact1} and \citet{LickHDFcompact2} observed 61
compact, high-surface-brightness galaxies ($r<0\farcs 5$) in the HDF-N
flanking fields.  The great majority show emission lines and have
redshifts $0.4<z<1$.  Masses for the systems were deduced from the line
widths, which ranged from $35-150 \kms$, and one-half of the sample
were found to be low mass ($M<10^{10} M_\odot$), relatively luminous
systems similar to local H II galaxies.  The remaining systems were a
heterogeneous class similar to local starburst galaxies.

\citet{VogtHDF} studied a 
separate magnitude-limited sample of eight high-inclination disk
galaxies from the HDF-N flanking fields in the redshift range
$0.15<z<0.75$.  Reliable velocity information could be
discerned to three disk scale-lengths, comparable to the extent of optical rotation
curves for local galaxies. Results from this survey were combined 
with an earlier survey of more luminous galaxies \citep{VFPGFIK96} to
study the Tully-Fisher (TF) relation at $z \sim 0.5$ over a span of 3 magnitudes.
A luminosity--line-width relation exists for the sample, 
linear and of the same slope as the local TF relation, 
but shifted to higher luminosities by 0.4 magnitudes. This is similar
to the results from other studies \citep{FPKI96,Bershady97p537}
but larger amounts of luminosity evolution were seen in other samples
\citep{RGCI97,SP98}. \citet{VogtHDF} speculate that
the differences are due to sample selection. In any case, the modest 
luminosity evolution is consistent with the modest surface-brightness
evolution found in the analysis of structural parameters by \citet{SKFSVPGIW99}.

\subsection{Mid-IR Sources} \label{sec_midIR}

Mid-IR observations offer the opportunity to detect emission from 
material heated by star formation or active galactic nuclear (AGN) 
activity, and thus provide some 
indication of radiative activity that may be obscured at UV through 
near-IR wavelengths.   The launch of the {\it Infrared Space Observatory} (ISO), 
the first space-based mid-IR telescope capable of such observations, coincided 
nicely with the availability of the HDF data set, and both deep fields (N and S) 
were the targets of deep 6.75$\mu$m and 15$\mu$m observations with the ISOCAM 
imager.   The HDF-S ISO observations were made before the HST imaging was 
carried out, and they are currently being analyzed (e.g., \citealt{Oetal00}).

The HDF-N ISO observations were carried out using discretionary time 
allocated by the ISO director to an observing team led by Rowan-Robinson
\citep{ISOHDF1,ISOHDF2,ISOHDF3,ISOHDF4,ISOHDF5}.
Two other groups have 
subsequently processed and reanalyzed the ISO data using independent 
methods \citep{DPCPABC99,ACES99}.  The 15$\mu$m maps 
extend well beyond the central HDF to cover a portion of the flanking fields, 
and have $\approx 9\arcsec$ resolution.  The high-sensitivity region of 
the 6.7$\mu$m maps is smaller and roughly matched to the central HDF WFPC2 
area, with $\approx 4 \arcsec$ resolution.  

The depth of the ISO image mosaics varies over the field of view, and the 
data processing procedures are complex and the subject of continuing refinement.
Extensive simulations have been used by the three groups to calibrate source 
detection, completeness and reliability, and photometry.    Given these 
complexities, it is perhaps not unexpected that the groups have produced 
different source catalogs.   A simple, visual inspection of the images presented 
by \citet{ISOHDF1} or \citet{ACES99} demonstrates that 
the 15$\mu$m data is much more straightforward to interpret -- at least $\sim$20 
sources are readily visible to the eye, whereas the 6.7$\mu$m image has only a 
few ``obvious'' sources and much more low-level background structure.   
Indeed, the source catalogs from the three groups agree reasonably well at 15$\mu$m, 
commonly detecting most of the brighter sources ($\sim 20$ detections with 
$\gtrsim 200 \mu$Jy)  
with generally similar measured flux densities.  
Aussel et al and Desert et al, however, claim reliable source detection to 
fainter flux limits than do Goldschmidt et al, and thus have longer source 
lists:  The complete plus supplemental catalog of Aussel \etal lists 93 sources 
at 15$\mu$m.  At 6.75$\mu$m, the situation is more confused (perhaps literally).
All three groups define ``complete'' lists of  six or seven sources, but they are not 
all the same objects.   Differences in the assumptions about ISO astrometry and 
geometric distortion corrections may be partially responsible.  Goldschmidt et al. 
provide a list of 20 supplemental sources at 6.7$\mu$m, only a few of which are 
detected by the other groups.  There also appear to be systematic differences 
in the derived source fluxes at this wavelength.     Altogether, the 15$\mu$m 
source detections and fluxes appear to be well established, but only the few brightest 
$6.7-\mu$m detections are unambiguous, and the associated fluxes may be 
uncertain.

The optical counterparts of the HDF-N ISO sources are mostly brighter galaxies at
$0 < z < 1.2$, although occasionally the identifications may be confused because of
the 9\arcsec\ PSF at 15$\mu$m.  Many of the 15$\mu$m sources also correspond to 
radio sources (few or none, however, correspond to the SCUBA sources discussed
below).  Over this 
redshift range, the 15$\mu$m measurements primarily sample emission from the 
``unidentified emission bands,''  including the strong features at 
$6.2-8.6 \mu$m.   These features are considered a characteristic signature
of star-forming galaxies (see, e.g., \citealt{RSGLMT99}).  
At $z > 1.4$, they shift out of the 15$\mu$m ISO bandpass,  and therefore more 
distant objects should become much fainter.  Indeed none have been identified 
yet in the ISO/HDF.   Warm dust ($T > 150$K) may also contribute to the mid-IR 
emission, and some of the ISO sources are almost certainly AGN (e.g., HDF 2-251, 
a red early-type radio galaxy at $z=0.960$, whose mid-IR emission is very 
likely not powered by star formation). Most of the HDF/ISO galaxies 
do not appear to be AGN, however, and if powered by reprocessed stellar UV radiation
have obscured star formation rates 
that greatly exceed those derived from their UV/optical emission 
\citep{ISOHDF5}.  Assuming an M82-like spectrum 
to derive $k$-corrections, the average HDF/ISO source is roughly 10$\times$ 
more luminous than M82 at 15$\mu$m \citep{EACDFFHPS99p999}.  
\citet{Rowan-Robinson99}
estimates that roughly two thirds of the star-formation at $z \sim 0.5$
is in dust-enshrouded starbursts such as those detected by ISO.

\subsection{Sub-millimeter Sources}\label{sec_submm1}

The commissioning of the sub-millimeter bolometer array camera SCUBA on the JCMT has opened
a potentially revolutionary new window on the distant universe, permitting the
detection of dust-reradiated energy from starburst galaxies and AGN out to almost
arbitrarily large redshifts (thanks to the strongly negative $k$-correction at
850$\mu$m, where SCUBA is most sensitive).  A series of SCUBA surveys 
(e.g. \citealt{SIB97,BCSFTetal98} 
) have now resolved a source population that 
comprises a substantial fraction of the far-IR background measured by COBE
and that may account for a significant fraction of 
the global radiative energy density from galaxies.  One of
the first and deepest of such surveys was carried out by \citet{HSDRBetal98},
who observed the HDF for 50 hours, detecting five sources at 850$\mu$m to a 4.4$\sigma$
limit of 2~mJy.  Given the 15\arcsec\ SCUBA beam size and possible pointing uncertainties,
the optical identifications for many (if not all) of these five sources remain in
doubt, and some may be blends due to multiple objects.  The brightest source,
however, was subsequently pinpointed by a 1.3mm interferometric observation from
institute de radioastronomie millimetrique (IRAM) 
(\citealt{DNGGCetal99}).  Unfortunately, this accurate position 
(which coincides with a microjansky radio source) did not fully settle the identification 
of the optical counterpart, falling halfway between two HDF galaxies separated 
by 2\arcsec\ (and almost certainly at very different redshifts).    The second 
brightest SCUBA source, falling just outside the primary WFPC2 field, has no obvious 
counterpart in the flanking field WFPC2 images, nor in the NICMOS data of 
\citet{DickinsonJdrop}.

Taking advantage of the frequent (but not universal) association of
sub-millimeter sources with centimeter radio detections, \citet{BCR99p279} have used SCUBA to observe 14 radio sources in the HDF flanking
fields with optical and near-IR counterparts fainter than $I > 25$ and
$K > 21$, respectively. They surveyed roughly half of the flanking
field area to a 3$\sigma$ depth of 6~mJy, detecting five of the ``blank
field'' radio source targets in addition to two new radio-quiet sub-millimeter
sources; however, none of the optical/near-IR bright sources were
detected at 850$\mu$m.  For those sub-millimeter
sources with radio counterparts, photometric redshifts from the
850$\mu$m to 20 cm flux ratio \citep{CY99,BCR99p279,CB99}, 
suggest that most lie in the redshift range $1 < z <
3$. Sub-millimeter sources without radio counterparts may represent a higher
redshift tail.

The HDF-N has also been imaged at 450$\mu$m \citep{HSDRBetal98} and 2.8mm
\citep{WW97}.  No sources were detected at either wavelength,
which is not surprising given the flux densities of the 850$\mu$m detections.

\subsection{Radio sources}

The HDF-N has become one of the best-studied regions of the sky at radio
wavelengths, and the availability of extremely deep optical/IR imaging and 
extensive spectroscopy has made it arguably the most important survey for 
understanding the properties of ``average'' radio sources at the millijansky and 
microjansky level.  The VLA and MERLIN arrays have been used to map regions including
the HDF-N at 3.5 cm and 20cm wavelengths
\citep{FKRWP97,RKFWP98,Richards99Lett,MWRKRG99,Richards00}.
The combined 
MERLIN + VLA data have been used to make high-resolution maps of 91 HDF radio 
sources, permitting a detailed study of radio source extent, as well as 
pinpointing locations for a few ambiguous identifications and confirming 
the blank-field character of others.  Altogether, 16 radio sources are 
located in the central HDF-N WFPC2 field, and many more in the flanking fields.

The optical counterparts for most of the radio sources (72 of 92 in the
region with suitably deep imaging) are relatively bright galaxies 
($ \langle I \rangle \approx 22$) at $0.4 < z < 1$.  Apart from a few manifestly 
AGN powered objects, usually giant elliptical radio galaxies (including the $z=1.02$ 
galaxy HDF 4-752.1), most are disk galaxies, sometimes disturbed or interacting. 
For nearly all, the radio emission is resolved with a typical size $\sim 1\arcsec$,
reinforcing the evidence that it primarily arises from star formation.
However, about 20\% of the sources have very faint or optically 
invisible counterparts with $I \gtrsim 25$.  Many of these have $K$-band 
identifications with very red optical-IR colors \citep{Richards99Lett}.
\citet{WWCPSS99} have studied one red HDF radio source which
is apparently at $z = 4.4$ (from a single-line spectroscopic redshift).
Several have been detected in 850$\mu$m observations 
with SCUBA \citep{BCR99p279}.  These ``invisible''
radio sources appear to be a population discontinuous from the majority of
the optically brighter sources, and may be high-redshift, dust-obscured, 
vigorously star-forming systems.  Finally, two VLA sources in the HDF proper have no optical
counterparts to the limits of the WFPC2 data, and one (123646+621226) has 
no counterpart to the extremely faint limit of the NICMOS GTO HDF field.

The HDF-S was observed by the Australia Telescope National Facility (ATNF)
in 1998.
The first year's observations covered a 60\arcmin\ diameter
region at 20~cm detecting 240 sources down to 100$\mu$Jy (13 in the primary
WFPC2 field) and correspondingly smaller fields at 13, 6 and 3 cm.
These have now been augmented by substantially deeper observations which
are currently being analyzed (Norris et al.\ 2000), but which should reach 
$5\sigma$ sensitivities of 40$\mu$Jy at 13 and 6~cm.

\subsection{Active Galactic Nuclei}
The number-density of AGN at HDF depths is of great interest because
of its connection to the X-ray background and to the re-ionization of
the universe at high redshift. \citet{JM98} identified
12 possible candidates with colors and morphologies consistent with 
AGN at $z > 3.5$. A similar study by \citet{CKMO99} found no
candidates above $z = 3.5$ and an upper limit of 20 candidates at
lower $z$.  These relatively low  number densities support
the prevailing view that the UV emission from AGN at high
redshifts is insufficient to account for the ionization of the IGM,
and they also suggest that black holes do not form with constant efficiency
within cold-dark-matter halos \citep{HML99}.
Fainter AGN yet to be identified may lie within the HDF.
Based on the properties of many of the optical counterparts to 
faint ROSAT X-ray sources, \citet{AF97} estimate
that roughly 10\% of the galaxies in the HDF are likely to be 
X-ray luminous, narrow-lined AGN. 
However, few of the optically identified AGN candidates in the HDF-N are
detected in the x-rays by the Chandra observatory \citep{HBGSBTBBCFGLNS00}.

\subsection{Gravitational Lensing}
The depths of the HDFs have made them well suited for studies of
gravitational lensing.  As \citet{Blandford98p245} noted in
his review, based on a rule of thumb from the number of lensed quasars
and radio sources in other surveys, roughly 0.2 \% of the galaxies,
corresponding to five to ten sources, were expected to be strongly lensed
(producing multiple images) in each HDF field.  In addition, weak
lensing by large-scale structure and individual galaxies should slightly distort
most  of the galaxy images.
Initial visual inspection of the HDFs produced a number of
possible strong lensing candidates \citep{Hogglens,BBH99},
but none has yet been confirmed by spectroscopy.
Rather, a few of the brighter examples appear not to be lensed systems,
but only chance superpositions \citep{ZMD97}.  A few candidates remain
in each of the primary WFPC2 HDF fields \citep{BBH99},
but it will require spectroscopy of very faint sources to confirm their
nature.

An analysis of the
HDF-N images by \citet{ZMD97} concluded that there were
at most one to two strongly lensed sources in the entire field, although very
faint objects with small angular separation could have escaped
detection.  On this basis they suggested that the HDF data were not
compatible with a large cosmological constant.  \citet{CQM99} took this further by utilizing published photometric redshifts
for the galaxies in the HDF-N to calculate the expected number of
multiply imaged galaxies in the field.  They found that a limit of one
detectable strongly lensed source requires $\Omega_\Lambda - \Omega_M <
0.5$.  For $\Omega_M, \Omega_\Lambda, \Omega_{\rm tot} = 0.3, 0.7, 1$
they predict 2.7 multiply imaged galaxies. The lensing statistics are
thus, in spite of earlier expectations for a larger number of lenses, 
not in conflict with recent determinations of
$\Omega_\Lambda$ from high-redshift SN \citep{PAGKNetal99,RFCCDetal98}.

Weak lensing is manifested by a tangential shear in the image of the
more distant source, and the amount of ellipticity, or polarization $p$,
introduced by a typical closely-spaced galaxy pair is of the order of $p \sim 
10^{-2}$.  
It is possible to attempt a statistical detection by looking for the
effect superimposed on the intrinsic morphologies of the source galaxies
perpendicular to galaxy-galaxy lines of sight.
As a differential effect, weak-lensing distortion is far less
sensitive to the cosmological model than to the masses or surface
densities of the intervening deflector galaxies.
Using color and brightness as redshift indicators,
\citet{DAT96} defined a sample of 650 faint
background and 110 lens galaxies in the HDF-N and
reported a
$3 \sigma$ detection equivalent to $p=0.06$ for galaxy-lens pairs having a
separation of $2\arcsec$. This corresponds to an average galaxy mass of
$6 \times 10^{11} M_\odot$ inside 20 kpc, or an internal velocity dispersion of 
$185 \kms$ 
A similar analysis of HDF-N images was subsequently performed
by \citet{HGDK98}, who used photometric
redshifts to determine distances of lens and source galaxies, and who
discriminated between lensing galaxy types based on colors.  Limiting
their analysis to separations greater than $3\arcsec$, they succeeded in measuring
background shear at a 99\% confidence level, finding that
intermediate-redshift spiral galaxies follow the Tully-Fisher relation
but are 1 mag fainter than local spirals at fixed circular velocity.
Although the sense of the evolution is the opposite of that found in
the kinematical studies, given the uncertainties the lensing result is 
consistent with the modest luminosity-evolution observed
by \citet{VogtHDF} when the studies are compared using the same cosmology.
The lensing results are inconsistent with the larger luminosity evolution
found by \citet{RGCI97} and \citet{SP98}.

\subsection{Intergalactic Medium}

The HDF-S STIS field was
selected to contain the bright, moderate redshift quasar J2233-606.
Deep HST imaging of the field around the quasar during the HDF-S
campaign has provided detection and morphologies of numerous
galaxies near the line of sight.
Moderate- to high-resolution spectra have
been obtained of the quasar both from the ground and with HST covering
the wavelength regime 1140-8200 {\AA} \citep{Savaglio98,SDW98,OBCHWN99,FergusonHDFS}, whereas follow-up
observations from the ground have established the redshifts of some of
the brighter galaxies in the field \citep{TDPCW99}.  
Too few redshifts have yet been measured to allow detailed study of
the correspondence between galaxies and QSO absorbers. 
A tentative detection of Ly$\alpha$
emission surrounding the QSO was reported by 
\citet{BPCABF99}.

Using the STIS UV and groundbased optical spectra, 
\citet{SFBESBCKSWW99} 
found the number density of Ly$\alpha$ clouds with $\log
N_{\rm HI} > 14$ in the redshift interval $1.5 < z < 1.9$, to be higher
than that found in most previous studies, and saw no evidence for a
change in the Doppler parameters of the  Ly$\alpha$ lines with
redshift. The two-point correlation function of Ly$\alpha$ clouds shows
clear clustering on scales less than $300 \kms$, especially the higher
column density systems, in agreement with results from other quasars.
Metal abundances in several absorption line systems have been studied
by \citet{PB99}. 
\citet{PS99}
studied the
metal lines in the absorption-line systems near the QSO redshift and
found that relative line strengths of different species are best
modeled with a multi-zone partial-covering model wherein different
species (e.g Ne VIII, O VI) arise from gas at different distances from
the AGN, and the clouds cover the central continuum emission region
completely but only a fraction of the broad emission-line region.

\subsection{Galaxy Clustering}\label{sec_clustering_phenomenology}


The measurements of clustering using the HDF have varied in 
the catalog and object selection, in the angular scales considered, in
the use or non-use of photometric redshifts, and in the attention
to object masking and sensitivity variations. 
In an early paper on the HDF-N, \citet{CROS96} explored whether the
galaxy counts in the HDF are ``whole numbers,''  i.e. whether fragments
of galaxies were incorrectly being counted as individual galaxies.
From an analysis of the
angular correlation function of objects with color-redshifts $z > 2.4$
they conclude that many of these objects are HII regions within a larger
underlying galaxy.  This analysis showed a strong clustering signal
on scales $0\farcs 2 < \theta < 10\arcsec.$ \citet{CGOR97} 
explored various hypotheses for the neighboring galaxies and
favored a scenario in which the faint compact sources in the HDF are
giant star-forming regions within small Magellanic irregulars.

It is not clear
whether the Colley et al. (1996, 1997) results apply to other catalogs,
given the different cataloging algorithms used. In particular, programs
such as SExtractor \citep{BA96}
and FOCAS \citep{TJ79}
use sophisticated and (fortunately or
unfortunately) highly tunable algorithms for merging or splitting
objects within a hierarchy of isophotal thresholds. The DAOFIND
algorithm used by \citet{CROS96} provides no such post-detection
processing. The extent to which this affects the results can only be
determined by an object-by-object comparison of the catalogs, which has
not (yet) been done in any detail. \citet{fergHDFsymp} (Figs. 1-4)
presents a qualitative comparison of several catalogs (not including
that of Colley et al.) which shows significant differences in how
objects with overlapping isophotes are counted. 
The \citet{CGOR97} analysis focused specifically on 695 galaxies
with color-redshifts $z > 2.4$. This number of galaxies is considerably
larger than the 69 identified by \citet{MFDGSF96}, using very conservative
color selection criteria, or the 187 identified by \citet{Dickinson98p219}
with somewhat less conservative criteria. For these smaller samples, 
we suspect that the ``overcounting'' problem is not as severe as
Colley et al. contend. 

Problems separating overlapping objects, although important
for clustering studies on small angular scales, do not have 
much effect on the overall galaxy number--magnitude relation \citep{fergHDFsymp} or angular
clustering measurements on scales larger than $\sim 2\arcsec$.

Other studies of clustering in the HDF have been restricted
to separations larger than $2\arcsec$.  The analysis has focused
primarily on the angular correlation function $\omega(\theta)$, 
which gives the excess probability $\delta P$, with respect to a random
Poisson distribution of $n$ sources, of finding two sources in solid angles
$\delta \Omega_1$, $\delta \Omega_2$ separated by angle $\theta$:
\begin{equation}
\delta P = n^2 \delta \Omega_1 \delta \Omega_2 [ 1 + \omega(\theta) ] .
\end{equation}
The angular correlation function is normally modeled as a power law 
\begin{equation}
\omega(\theta) = A(\theta_0) \theta ^ {1-\gamma}.
\end{equation}
Because of the small angular size of the field, 
$\omega(\theta)$ is suppressed if the integral of the correlation
function over the survey area is forced to be zero. Most authors
account for this ``integral constraint'' by including another parameter
and fitting for $\omega(\theta) = A(\theta_0) \theta ^ {1-\gamma} - C$,
with a fixed value of $\gamma = 1.8$ and with $A$ and $C$ as 
free parameters.

\citet{VFdaC97} measured the overall
angular correlation function for galaxies brighter than $R = 29$
and found an amplitude $A(\theta = 1\arcsec)$ decreasing with increasing
apparent magnitude, roughly consistent with the extrapolation of
previous ground-based results. The measured amplitude was roughly
the same for the full sample and for a subsample of red galaxies
(considered typically to be at higher redshift). The motivation
for this color cut was to try to isolate the effects of magnification
bias due to weak lensing by cosmological large-scale structure
\citep{MJV98}, but the predicted effect is small and could
not be detected in the HDF (and will ultimately be difficult
to disentangle from the evolutionary effects discussed in 
\S\ref{sec_clustering}). 

The remaining HDF studies have explicitly used photometric redshifts.
\citet{CSB99} considered scales $3\arcsec < \theta < 220\arcsec$ and
galaxies brighter than $I_{814} = 27$.  Within intervals $\Delta z =
0.4$ the amplitude $A(\theta = 10\arcsec) \sim 0.13$ shows little sign of
evolution out to $\zphot = 1.6$ (the highest considered by Connolly et
al.).  \citet{RVMB99} analyzed a $U$-band selected sample,
isolated to lie within the range $1.5 < \zphot < 2.5$ and angular
separations $2\arcsec < \theta < 40\arcsec$. The results are consistent with those
of \citet{CSB99}, although \citet{RVMB99} point out that
galaxy masking and treatment of the integral constraint can have a
non-negligible effect on the result. 

Measurements of $\omega(\theta)$ for samples extending out to 
$\zphot > 4$ have been carried out by \citet{MagMad99}, 
\citet{ACMMLFG99} and \citet{MP98}. The minimum
angular separations considered for the three studies were
$9\arcsec$, $5\arcsec$, and $10\arcsec,$ respectively: i.e. basically disjoint from
the angular scales considered by \citet{CGOR97}.
All three studies detect increased clustering at $z \gtrsim 2$,
although direct comparison is difficult because of the
different redshift binnings. The interpretation shared by 
all three studies is that the HDF shows strong clustering
for galaxies with $z \gtrsim 3$, in qualitative agreement
with Lyman-break galaxy studies from ground-based samples
\citep{GSADPK98,ASGDPK98}. Although this may be the correct
interpretation, the significant differences in
the photometric redshift distributions and the derived clustering
parameters from the three studies leave a considerable
uncertainty about the exact value of the clustering amplitude.
The two most comprehensive studies differ by more than
a factor of 3 in $A(\theta=10\arcsec)$ at $z = 3$ \citep{ACMMLFG99,MagMad99}.
Both studies use photometric redshifts 
to magnitudes $I_{814} = 28$, which is fainter
than the detection limits in the F300W and F450W bands for even
flat spectrum galaxies. A large part of the disagreement may 
thus be due to scatter in $\zphot$. The analysis also hinges
critically on the assumed power-law index $\gamma = 1.8$ and 
the necessity to fit the integral constraint. Thus, although it seems
reasonably secure that a positive clustering signal has been measured
in the HDF at high $\zphot$, it will require much larger data sets to
constrain the exact nature of this clustering
and determine its relation to clustering at lower redshift.

At brighter magnitudes, spectroscopic surveys
show clear evidence for clustering in redshift space
\citep{CCHSBHS96,ASGDPK98,Cohen99}, with pronounced peaks even
at redshifts $z > 1$.
These structures, and others within the flanking fields,
do not show evidence for centrally concentrated structures, and are
probably analagous to walls and filaments observed locally.

\section{Interpretation}
\subsection{Galaxy counts vs. simple models}\label{sec_counts2}

The time-honored method of comparing cosmological models to field-galaxy
surveys has been through the classical number--magnitude, size--magnitude,
magnitude--redshift, etc. relations
\citep{Sandage88}. Much of this effort, including early
results from the HDF, has been reviewed in \citet{Ellis97}.
There are three important changes to the scientific
landscape that we must acknowledge before proceeding to discuss the more
recent interpretation of the HDF number counts. The first change is the transition
(motivated by high-$z$ SNe Ia, cluster baryon fractions, etc.)
from a favored cosmology with $\Omega_M = \Omega_{\rm tot} = 
1$ to one with $\Omega_M,\Omega_\Lambda,\Omega_{\rm tot} = 0.3,0.7,1$.
The latter cosmology comes closer to matching the galaxy counts
without extreme amounts of evolution. The second change is the realization that 
galaxy counts at HDF depths push to depths where the {\it details}
of galaxy formation become extremely important. Models that
posit a single ``epoch of galaxy formation'' were never realistic, and
are no longer very useful. Finally, it has become more
popular to model quantities such as the metal enrichment history $\MER$,
than it is to model galaxy counts. Consequently, to our 
knowledge there are no published papers that compare the 
overall $N(m)$ predictions of hierarchical semi-analytic models 
with the currently favored cosmology to the HDF galaxy counts.

\subsubsection{No-evolution (NE) models}

The assumption of no evolution is not physically reasonable, but
provides a useful fiducial for identifying how much and what kinds of
evolution are required to match faint-galaxy data. Traditional
no-evolution models are based on estimates of the $z=0$ luminosity
functions for different types of galaxies.  \citet{BBS97}
construct a non-evolving model from the HDF itself, using 32 galaxies
brighter than $\I = 22.3$ to define a fiducial sample. They construct
Monte-Carlo realizations of the HDF that might be seen from a universe
uniformly populated with such galaxies, shifted to different redshifts
and $k$-corrected on a pixel-by-pixel basis. This kind of simulation
automatically incorporates the selection and measurement biases of the
HDF at faint magnitudes.  It also normalizes the model {\it by fiat} to
match the counts at $\I = 22.3$, where traditional no-evolution models,
normalized to the local luminosity function, already see significant
discrepancies for the Einstein-de Sitter model.
The resulting models underpredict the HDF counts at $\I
= 27$ by factors of 4 and 7 for models with $\Omega_M = 0.1$ and 1.0
(with $\Omega_\Lambda = 0$), respectively. The angular sizes of
galaxies in the models are also too big at faint magnitudes, with a
median half-light radius about a factor of 1.5 larger than that observed for
galaxies with $24 < I < 27.5$.  Because the typical redshift of the
template galaxies in this model is $z \sim 0.5$, this model comparison
suggests that much of the evolution in galaxy number densities, sizes,
and luminosities occurs at higher redshift.  The typical small sizes of
faint galaxies essentially rule out low-surface-brightness galaxies
\citep{FM95,MR95} as a significant contributor to the counts at
magnitudes $\I > 20$ \citep{Ferguson_photz}.

\subsubsection{Pure-Luminosity Evolution models}

Many of the models that have been compared to the HDF number counts are
variants of {\it pure luminosity evolution} (PLE) models \citep{Tinsley78},
wherein galaxies form at some redshift $z_f$, perhaps varying by type,
with some star-formation history $\psi(t)$. There is no merging.
\citet{MSCFG96} compare several different models to the counts
and colors of galaxies in the HDF and in deep images taken at the
William Herschel Telescope. For low $\Omega$ a reasonable fit to $I \approx 26$
is achieved, but the model progressively underpredicts the counts to
fainter magnitudes, and the long star-formation timescales and heavily
dwarf-dominated IMF adopted for ellipticals in this model seems inconsistent
with the fossil evidence in local ellipticals.
By including a simple prescription
for dust attenuation, \citet{CS97} are able to achieve a 
reasonable fit to the counts for low $\Omega_M$ without resorting 
to a peculiar IMF.
Another set of PLE models was considered by \citet{PMZFB98}, with 
emphasis on the near-UV counts and the effect of UV
attenuation by intergalactic neutral hydrogen.  From color-magnitude relations and a
study of the fluctuations in the counts in the different HDF bands,
\citet{PMZFB98} conclude that at $\B = 27$ roughly 30\% of the sources
in the HDF are at $z > 2$. The PLE model considered has $\Omega_M =
0.1$ with no cosmological constant, and with a redshift of formation
$z_f = 6.3$. This model matches the counts quite well but predicts that
about 80 objects brighter than $\V = 28$ should disappear from the
F450W band because of their high redshift, although \citet{MFDGSF96} identify
only about 15 such sources.  \citet{FB98} considered
another low$-\Omega$ PLE model and encountered similar problems, predicting
roughly 400 B-band Lyman-break objects where only 15 or so were observed.
The utility of PLE models clearly breaks down for $z \gtrsim 2$,
where the details of galaxy formation become critical.

\subsubsection{Models with Additional Galaxy Populations}

The great difficulty in achieving a fit with $\Omega_M = \Omega_{\rm
tot}= 1$, even to ground-based galaxy counts, motivated investigations
into different kinds of galaxies that might be missed from the census
of the local universe but could contribute to the counts of galaxies at
faint magnitudes \citep{FM95,BF96,KGB93,MR95}.
Perhaps the most physically motivated of these more exotic
possibilities is the idea that the formation of stars in low-mass
galaxy halos could be inhibited until low redshifts $z \lesssim 1$ because
of photoionization by the metagalactic UV radiation field
\citep{Efstathiou92,BR92}.  \citet{FB98} compared the
predictions of such an $\Omega_M=1$ ``disappearing dwarf'' model in
detail to the HDF. They found that the simplest version of the model
({\it a}) overpredicts the counts at faint magnitudes, and ({\it b}) overpredicts
the sizes of very faint galaxies. These problems are caused by the fact
that, for a Salpeter IMF, the dwarfs fade too slowly and would still be
visible in great numbers in the HDF at redshifts $z < 0.5$.
\citet{Campos97} considered a model with much milder evolution, with
each dwarf undergoing a series of relatively long (a few $\times 10^8$
yr) star-formation episodes. Acceptable fits to the counts and colors
of galaxies are achieved both for high and low values of $\Omega$. Both
the \citet{FB98} and the \citet{Campos97} models predict that the
HDF sample at $I > 25$ is dominated by galaxies with $z < 1$, a result
inconsistent with existing photometric-redshift measurements.  Using a
volume-limited photometric redshift sample to construct the bivariate
brightness distribution of galaxies with $0.3 < z < 0.5$, \citet{Driver99} concludes that the volume density of low-luminosity,
low-surface-brightness galaxies is not sufficient to explain the
faint-blue excess either by themselves or as faded remnants.  
Further constraints on the low-redshift, low-luminsity population can be
expected from the HDF STIS UV observations.

\subsection{The morphology of high-redshift galaxies}\label{sec_morph2}

As was described in \S\ref{sec_morph1}, HST images from the HDF and other
surveys have established that the fraction of irregular and peculiar galaxies
increases toward faint magnitudes (down to $I \sim 25$, where galaxies become
too small for reliable classification).
Some of the morphological peculiarities in HDF galaxies may be the
consequence of interactions, collisions, or mergers.
A higher merger rate
at earlier epochs is a natural consequence of models that assemble
the present-day Hubble sequence galaxies by a process of hierarchical
mergers (\citealt{BCF96}).  In the local universe, morphological
asymmetry correlates reasonably well with color, with late type, blue
galaxies and interacting objects 
showing the greatest asymmetry (\citealt{CBJ99}).  In the
HDF, however, there are also highly asymmetric galaxies with {\it red}
rest-frame $B-V$ colors as well as blue ones, and the asymmetries
persist at NICMOS wavelengths (cf. Fig. \ref{fig_morphology}), where
longer-lived stars in a mixed-age stellar population would dominate
the light and where the obscuring effects of dust would be reduced.
The timescale over which an interacting galaxy will relax and regularize
should be shorter ($\lesssim 1$~Gyr) than the time over which stars
formed during the interaction would burn off the main sequence, and thus
if star formation occurs during collisions then blue colors should persist
longer than the most extreme manifestations of morphological disturbance.
\citet{CBDFPetal00} suggest that the large number of asymmetric HDF
galaxies demonstrates the prevalence of early interactions and mergers,
some of which (the bluest objects) have experienced substantial star
formation during the encounter, whereas others have not (or, alternatively,
have their recent star formation obscured by dust).

Although interactions seem an attractive way of explaining these
morphological peculiarities, it is nevertheless worth mentioning some
qualifications concerning the details of the models which have been tested
to date.  First, although a high merger rate at relatively low redshifts is
a generic prediction of the standard cold dark matter (SCDM) model, 
models with lower $\Omega_M$ have a substantially lower merger rate
at $z < 1$.  The \citet{BCF96} model that was explicitly compared with
the HDF was based on the SCDM cosmology.
Second, even with the high merger rate the 
\citet{BCF96} model predicts that galaxies that have had major merger within 1
Gyr prior to the time of observation constitute less
than 10\% of the total population at $I = 25$ (their Fig. 2).   Most of
the irregular galaxies in this model are simply bulgeless, late-type
galaxies.  \citet{IGNRRGS99} have studied the redshift distribution of
galaxies with $17 < I < 21.5$ and conclude that the irregular/peculiar class is a
mix of low-redshift dwarf galaxies and higher-redshift ($0.4 < z < 1$)
more luminous galaxies.  These higher redshift galaxies are unlikely to
be the progenitors of present-day dwarf irregular galaxies, but it is
not clear whether they are predominantly low-mass galaxies undergoing
a starburst, or more massive galaxies undergoing mergers.  The scatter
in colors and sizes suggests it is a mix of both, but more detailed
kinematical information is needed.

\subsection{Elliptical galaxies}\label{sec_ellipticals}

The simple picture of the passively evolving elliptical galaxy, formed
at high redshift in a single rapid collapse and starburst, has held sway
since the scenario was postulated by \citet{ELS62} and its photometric consequences were modeled by \citet{Larson74} and \citet{TG76}. This hypothesis is supported by
the broad homogeneity of giant elliptical (gE) galaxy photometric and
structural properties in the nearby universe, and by the relatively
tight correlations between elliptical galaxy chemical abundances and
mass or luminosity.  Observations in the past decade have pushed toward
ever higher redshifts, offering the opportunity to directly watch the
evolutionary history of elliptical galaxies.  Most of this work has
concentrated on rich cluster environments, and is not reviewed here
except to note that most observers have favored the broad interpretation
of quiescent, nearly passive evolution among cluster ellipticals out to
$z \approx 1$ (e.g. \citealt{AECC93}; \citealt{SED95,SED98};
\citealt{ESDCOBS97};
\citealt{vDFKIFF98,DPSEDE99}).

The ``monolithic'' formation scenario serves as a rare example of a clearly
stated hypothesis for galaxy evolution against which to compare
detailed measurements and computations.  The alternative
``hierarchical'' hypothesis is that elliptical galaxies formed mostly
via mergers of comparable-mass galaxies that had at the time of
merging already converted at least some of their gas into stars (e.g.
\citealt{KWG93}). The argument between the monolithic and the
hierarchical camps is not about the origin of galaxies from
gravitational instability within a hierarchy of structure, but rather
about when gEs assemble most of their mass and whether
they form their stars mostly {\it in situ} or in smaller galaxies that
subsequently merge.  At least one observational distinction is clear:
In the hierarchical model the number density of elliptical galaxies
should decrease with redshift. In the monolithic model the number
density should remain constant and the bolometric luminosities should
increase out to the epoch of formation.

Measuring the density and luminosity evolution of the {\it field}
elliptical population, however, has proven difficult, and many different
approaches, perhaps complementary but not necessarily concordant,
have been used to define suitable galaxy samples.  Recent debate
has focused on estimates of the evolution the co-moving density of
passively-evolving elliptical galaxies in the Canada-France Redshift
Survey \cite{CFRS1}. Applying different photometric selection criteria and
different statistical tests, two groups \citep{CFRS6,TY98}
found no evidence for density evolution out to $z = 0.8$, whereas another
group \citet{KCW96}, found evidence for substantial
evolution.  This debate highlights the difficulties inherent to defining
samples based on a color cut in the margins of a distribution, where small
changes in the boundary, as well as systematic and even random photometric
errors in the data, can have substantial consequences for the conclusions.

HST, makes it possible to define samples
of distant galaxies morphologically.  The multicolor HDF images provide an
attractive place to study distant ellipticals, but because of its very small
volume and the inherently strong clustering of elliptical galaxies,
one must be careful in drawing sweeping conclusions from HDF data alone.
Thus, although we focus our attention primarily on the HDF
results, they should be considered in context with results from other surveys
(e.g. \citealt{DWG95,GESG95,DFCOWPLY98,TSCMB99,TS99}).

The interpretation of the high-redshift elliptical galaxy counts
depends in large measure on a comparison to the {\it local} luminosity
function (LF) of elliptical galaxies.  The basic parameters of
published LFs for local elliptical galaxies (or what are
sometimes assumed to be elliptical galaxies)\footnote{even locally,
classification uncertainties may be partly responsible for the widely
diverse measurements of the elliptical galaxy LF (see \citealt{LPEM92,MGHC94,MdCPWG98,ZPZ94,Sandage00}).} span a very wide
range in normalization, characteristic luminosity, and faint end behavior. 
Early HST studies such as that by \citet{IGRS96} concluded
that NE or PLE models were consistent with elliptical-galaxy number counts
predicted using the local LF of \citet{MGHC94}.

Searches for distant elliptical galaxies in the HDFs have relied on either color or
morphology. In the HDF-N, \citet{FCAF98} and \cite{FF98}, along with \citet{Schade99} for the CFRS and LDSS surveys, 
all examined the size-luminosity relation for morphologically-selected
ellipticals, finding evolution out to $z \sim 1$ consistent with PLE models.
\citet{KBB99} found a well-defined color-magnitude
sequence at $\langle z \rangle \sim 0.9$, consistent with passive
evolution for approximately half the galaxies, but they also noted a
substantial ``tail'' of bluer objects. Similarly, in the 
\citet{Schade99} study, about one third of the sample at $z > 0.5$ 
had [OII] line emission and colors significantly bluer than PLE models.

Ir imaging has made it attractive to
pursue gEs at redshifts $z > 1$.  In the HDF-N, \citet{Zepf97} and
\citet{FSFGBAD98} used ground-based $K$-band data, selecting samples
by color and morphology, respectively, and concluded that there was an absence of
the very red galaxies that would be expected if the elliptical population as
a whole formed at very large redshift and evolved passively.  In particular,
Franceschini et al.\ highlighted the apparently sudden disappearance of HDF
ellipticals beyond $z > 1.3$, which suggests that either dust obscuration during
early star formation, or morphological perturbation during early mergers,
was responsible.  \citet{BCTFHSH99} made an 
IR study of the HDF-N flanking fields, selecting objects by
color without reference to morphology.  To $K \approx 20$, they found few
galaxies with $I - K > 4$, the expected color threshold for old ellipticals
at $z \gtrsim 1$.    However, other comparably deep and wide IR
surveys have reported substantially larger surface densities of
$I - K > 4$ galaxies (\citealt{EESDSSD98,MMSCGF00}),
raising concerns about field-to-field variations.  \citet{MEABC99}
studied the optical-to-IR color distribution for a sample of $\sim 300$
morphologically early-type galaxies selected from 48 WFPC2 fields,
including the HDF-N and flanking fields.  They too find an absence of very red
objects and a generally poor agreement with predictions from purely
passive models with high formation redshifts, although the sky surface
density agrees reasonably well with the sorts of models that matched the
older MDS and HDF counts, i.e., those with a suitably tuned local LF.

On HST, NICMOS has provided new opportunities to identify and study
ellipticals at $z \gtrsim 1$.  Its small field of view, however, has limited
the solid angle surveyed.   Spectroscopic
confirmation of the very faint, very red elliptical candidates 
identified so far will be exceedingly difficult, but would be well
worth the effort.
\citet{TSWWBBBCCDetal98}, \citet{StiavelERO99} and
\citet{BBBSR99} noted several very red, $R^{1/4}$-law galaxies
in the HDF-S NICMOS field, identifying them as ellipticals at
$1.4 \lesssim z_{\rm phot} \lesssim 2$.  Given the small solid angle
of that image, this suggests a large space density, and Ben\'{\i}tez
et al.\ have proposed that most early type galaxies have therefore
evolved only passively since $z \sim 2$.   Comparably red, spheroidal
galaxies are found in the NICMOS map of the HDF-N
(\citealt{DickinsonHDFNIC}), with photometric redshifts in the range
$1.2 \lesssim z \lesssim 1.9$.   Their space density appears to be well
below that of HDF ellipticals with similar luminosity at $z < 1.1$,
in broad agreement with \citet{Zepf97} and \citet{FSFGBAD98},
although the comparison to $z=0$ is again limited primarily by uncertainties
in the local gE LF.  Curiously, few fainter objects with similar
colors are found in the HDF-N, although the depth of the NICMOS images
is adequate to detect red ellipticals with $L \sim 0.1 L^\ast$ out to
$z \approx 2$.  \citet{TS99} find that PLE models with 
high ($z \sim 5$) and low ($z \sim 2$) formation redshifts over- and 
under-predict the observed counts, respectively, of red elliptical-like objects
in 23 other NICMOS fields.

The HDF-N NICMOS images from \citet{DickinsonHDFNIC} and
\citet{ThompsonHDF} are deep enough to have detected red, evolved
elliptical galaxies out to at least $z \sim 3$ if they were present,
eliminating concerns about invisibility because of $k$-corrections (e.g.,
\citealt{Maoz97}).  The only plausible $z > 2$ candidate is an HDF-N
``$J$-dropout'' object, whose colors resemble those of a maximally
old gE at $z \sim 3$ \citep{LYF98,DickinsonJdrop,Lanzetta_photz}.
Other $z > 2$ HDF objects which
some authors have morphologically classified as ellipticals (cf.\
\citealt{FCAF98}) are blue, mostly very small, manifestly
star forming ``Lyman break'' objects.  Few if any appear to be passively
evolving objects that have ceased forming stars, so the connection
to present-day ellipticals is more speculative.

The collective evidence surveyed above suggests that mature, gE
galaxies have been present in the field since $z \sim 1$ with
space densities comparable to that at the present era.
At $0.4 < z < 1$, however, they exhibit an increasingly broad range
of colors and spectral properties, which suggests a variety of star
formation histories over the proceeding few billion years (or alternatively
errors in classification).
The statistics seem to favor a substantial decline in their space density
at $z \gg 1$, although 
the well-surveyed sightlines are small and few,
and clustering might (and in fact, apparently does) cause large variations
from field to field.  Therefore, the conclusion has not been firmly established.
Moreover, it is very difficult to achieve
uniform selection at all redshifts, regardless of the criteria used
(photometric, morphological, or both), and the existing samples of
objects with spectroscopic (or at least well-calibrated photometric)
redshifts are still small.  Thus even the deceptively simple task of
comparing elliptical galaxy evolution to the simple PLE hypothesis 
remains a stubborn challenge.

\subsection{Obscured populations}

ISO and SCUBA observations of the HDF and other fields have 
revealed an energetically important population of dust-obscured 
objects. Interpretation of these results has been the subject of considerable
debate, because of ambiguities in source identification and in distinguishing
starbursts from AGN. 
The ISOCAM (6.7 and 15 $\mu$m) and SCUBA (850 $\mu$m) observations 
bracket but do not sample the wavelength regime $100 <
\lambda_p < 200 \,\mu$m near the peak of the far-IR emission, making it
difficult to assess reliably the source contribution to the global
emissive energy budget of galaxies.  Although strong mid-IR emission
accompanies vigorous star formation in many nearby galaxies, the
unidentified IR emission bands carry most of the energy in the
wavelength range sampled by ISOCAM at $z \sim 1$. The bulk of
re-emitted radiation, however, emerges near $\lambda_p$, and there is
considerable diversity in $f(10 \mu{\rm m}) / f(100 \mu{\rm m})$ flux ratios
among nearby luminous and ultraluminous IR galaxies.  Therefore,
deriving star formation rates from mid-IR measurements alone requires a
substantial extrapolation and is quite uncertain.   SCUBA 850$\mu$m
observations sample the re-radiated thermal emission directly, but at a
wavelength well past $\lambda_p$, again requiring an extrapolation to
total far-IR luminosities assuming a dust temperature and emissivity
that are almost never well constrained by actual,
multi-wavelength measurements.  Furthermore, SCUBA and
ISO do not in general detect the same sources. ISO detects objects
out to $z \sim 1$, whereas most of the SCUBA sources could be at much higher redshifts.
For both mid-IR and sub-millimeter sources,
AGN-heated dust may also play a role.  Several active galaxies
(including radio ellipticals) in the HDF are detected by ISO, and some
SCUBA sources (not yet in the HDF, however) have been identified with
AGN. Chandra x-ray observations \citep{HBGSBTBBCFGLNS00} do not detect
with high significance any of the sub-millimeter sources in the HDF.

Difficulties in identifying the optical counterparts to the 
mid-IR and sub-millimeter sources are a second source of ambiguity.
The mean separation between galaxies in the HST images of the HDF
is about $3 \arcsec$. In comparison
the ISO 15$\mu$m PSF has full-width at half max (FWHM)~$\approx 9\arcsec$, whereas 
the SCUBA 850$\mu$m
beam size is $\approx 15\arcsec$.  Nevertheless, the HDF mid-IR sources have 
plausible counterparts among the brighter galaxies in the survey.  SCUBA
sources, on the contrary, often seem to have have extremely faint, and sometimes 
entirely invisible, counterparts in the optical and near-infrared.  Sub-mm
objects sometimes correspond to microjansky radio sources at centimeter wavelengths, and
sometimes have faint, very red near-IR counterparts.  Many, however,
do not, even with the deepest radio and near-IR data available with current
instrumentation (see \S\ref{sec_submm1}).

Other than the occasional ultra-red optical counterpart (of which
there are none in the HDF-N), the non-AGN counterparts to sub-millimeter and
far-IR sources rarely have particularly unusual photometric or spectral
features which highlight them as the remarkable objects they must be.
Counterparts to $6.7$ and 15$\mu$m ISO sources show a predominance of
``post-starburst'' Balmer absorption spectra (\citealt{FloresAA99,FloresApJ99,AECS99p1023,CHBCHSS00}); apart from the occasional AGN, few
have strong emission line spectra, which suggests the most intense
star-forming regions are highly obscured. 
There are hints that the Lyman-break galaxies represent the
faint tail of the sub-millimeter source population. \citet{CSSBHMADGP99}
detect one out of 18 galaxies in a targeted study of
$z \sim 3$ Lyman-break galaxies. \citet{PRBDEetal99} find a
significant statistical correlation of Lyman-break galaxies with
sky fluctuations in the HDF-N SCUBA map. A straightforward analysis
suggests that the ratio of hidden star formation to star formation
directly measured in the UV is about 6:1, and that Lyman-break
galaxies account for at least 25\% of the 850$\,\mu \rm m$ background.

Given the present status of faint sub-millimeter and mid-IR surveys, the most 
telling information comes not from individual source identifications but 
from the ensemble statistics using the combined data from many surveys.
The ISO 15$\mu$m number counts show good agreement between various surveys 
and a strong excess over no-evolution models (\citealt{EACDFFHPS99p999}), 
as do 850$\mu$m SCUBA counts 
(\citealt{BCS99};  \citealt{BKIS99}).  
As with the ``faint blue galaxy'' 
excess \citep{Kron80}
that helped spark the boom in the optical study of distant field galaxies 
in the 1970s and 1980s, the ISO and SCUBA counts point toward strong 
cosmological evolution, probably manifesting the star formation history of 
the galaxy population.  And like the faint blue galaxy problem, the robust 
interpretation of this number count excess will undoubtedly require extensive 
follow-up observations to characterize the source population and its 
properties.   Currently, given the uncertainties in source identification, 
in extrapolating the current measurements to ``true'' far-IR luminosities,
and in balancing the roles of AGN vs.\ star formation, it seems premature 
to attempt a detailed revision of our picture of galaxy evolution and 
cosmic star formation.  But it is also abundantly clear that a true 
understanding will have to account for the important and perhaps 
dominant energetic role played by this obscured population. 

\subsection{Global star-formation history and chemical evolution}\label{sec_lumden}

Interest in the integrated background light from galaxy formation is
long-standing (e.g. \citealt{BGH55}), and has motivated a large number
of experiments aimed at measuring the diffuse extragalactic background
(e.g. \citealt{SS78,DWW79,MAM88,Bernstein97,Hauser_etal98}).  Although the integrated background records the light from all
galaxies, whether or not they are individually detected, the task of
separating out galactic foregrounds and instrumental backgrounds is
formidable, and many of the measurements are only upper limits.  An
alternative approach is to add up the UV emission from galaxies that
are individually detected. This approach formally produces only a lower
limit for the true UV luminosity density, but it provides a basis for
exploring connections of the high- and low-redshift universe and for
deciding which of the various selection effects are plausible and
important.

The UV emission from galaxies is directly connected to their metal
production \citep{CLGM88} because the UV photons come from the same
massive stars that produce most of the metals through type II SNe.  The
relation between UV emissivity and metal production is not strongly
dependent on the initial mass function, varying by a factor of only 3.3
between the \citet{Salpeter55} and \citet{Scalo86} forms. In
contrast, the relation of UV emission to the total star-formation rate
is much more tenuous because the low-mass end of the IMF contains most
of the mass whereas the high-mass end produces most of the UV emission.
As a result, it is easier to constrain the history of metal production
in the universe than it is to constrain the overall star-formation
history. Prior to the HDF, estimates of the metal-production vs. redshift, 
$\MER$ had been
made from ground-based galaxy redshift surveys \citep{CFRSletter} 
and from QSO absorption line statistics (e.g.
\citealt{LWT95,PF95,FCP96}), and the observations at the time
suggested an order-of-magnitude increase in $\MER$ from $z=0$
to $z=1$.

\citet{MFDGSF96} made the first attempt to connect the luminosity density
in HDF high-redshift galaxy samples to lower-redshift surveys.
The plot of the metal-formation rate vs. redshift
has provided a focal point for discussion
of the HDF and for comparisons to theoretical models.
In the \citet{MFDGSF96} paper, galaxies in redshift slices
$2 < z < 3.5$ and $3.5 < z < 4.5$ were identified by strict 
color-selection criteria that were shown via simulations 
to provide strong rejection of objects outside of the desired
redshift intervals. The luminosity of the galaxies within these
redshift intervals was estimated by simply summing the observed fluxes
of the galaxies, and no dust or surface-brightness corrections were applied. 
The results were presented as lower limits because nearly all
of the corrections will drive the derived metal-formation rates
up. The initial \citet{MFDGSF96} diagram showed a metal-production
rate at $z \sim 4$ that was roughly a factor of 10 lower than the rate
at $z \sim 1$. \cite{MadauMaryland} subsequently modified the
color-selection criteria and integrated down an assumed luminosity
function, revising the $z > 2$ rates upward by about a factor of 3.
There was remarkable agreement between $\MER$ derived
from the galaxy luminosities, the results
of \citet{PF95}, and the predictions of hierarchical models
(e.g. \citealt{WF91,CAFNZ94,BCFL98}), all of which
show a peak in the metal production rate at $z \sim 1-2$.
Subsequent work in this area has focused on ({\it a}) galaxy
selection ({\it b}) effects of dust, and ({\it c}) the connection of
the Madau diagram to general issues of galaxy evolution and
cosmic chemical evolution, which we discuss in turn.

\subsubsection{Galaxy Selection}

The color-selection criteria of \citet{MFDGSF96} are 
extremely conservative, and
spectroscopic surveys \citep{SGDA96,LickHDFhighz} have identified at
least a dozen $2 < z < 3.5$ galaxies with $U_{300} - B_{450} > 1.3$ but with $B_{450}
- I_{814}$ redder than the \citet{MFDGSF96} selection boundary.
Alternative color-selection criteria (particularly in the $2 < z <
3.5$ range) have been explored in a number of studies
\citep{CC96,LickHDFhighz,MPD98,MHC99}, with the 
result that the luminosity-density goes up as the area in color space
is enlarged. The number of low$-z$ interlopers also may go up, 
although spectroscopic surveys
suggest that the \citet{MHC99} selection boundary is not prone
to this problem.
Similar techniques have been applied to ground-based images
covering much wider areas but sensitive only to brighter objects (see,
e.g. \citealt{SAGDP99} and \citealt{GDFCEHM98} for recent
examples).  Substantial progress has been made both in refining the
estimates of the volume sampled by the color selection and in estimates
of the high$-z$ galaxy luminosity functions \citep{Dickinson98p219,SAGDP99}.  These allow more precise (but more
model-dependent) estimates of the luminosity density at $z>3$.
In Fig. \ref{fig_lumden} we provide an updated plot of star-formation
vs. redshift from the color-selected samples of \citet{SAGDP99} and \cite{CasertanoHDFS},
now including integration down the luminosity function and a 
correction for mean dust attenuation. 


\ifsubmode\else

\begin{figure}
\ifsubmode\else
\centerline{\plotone{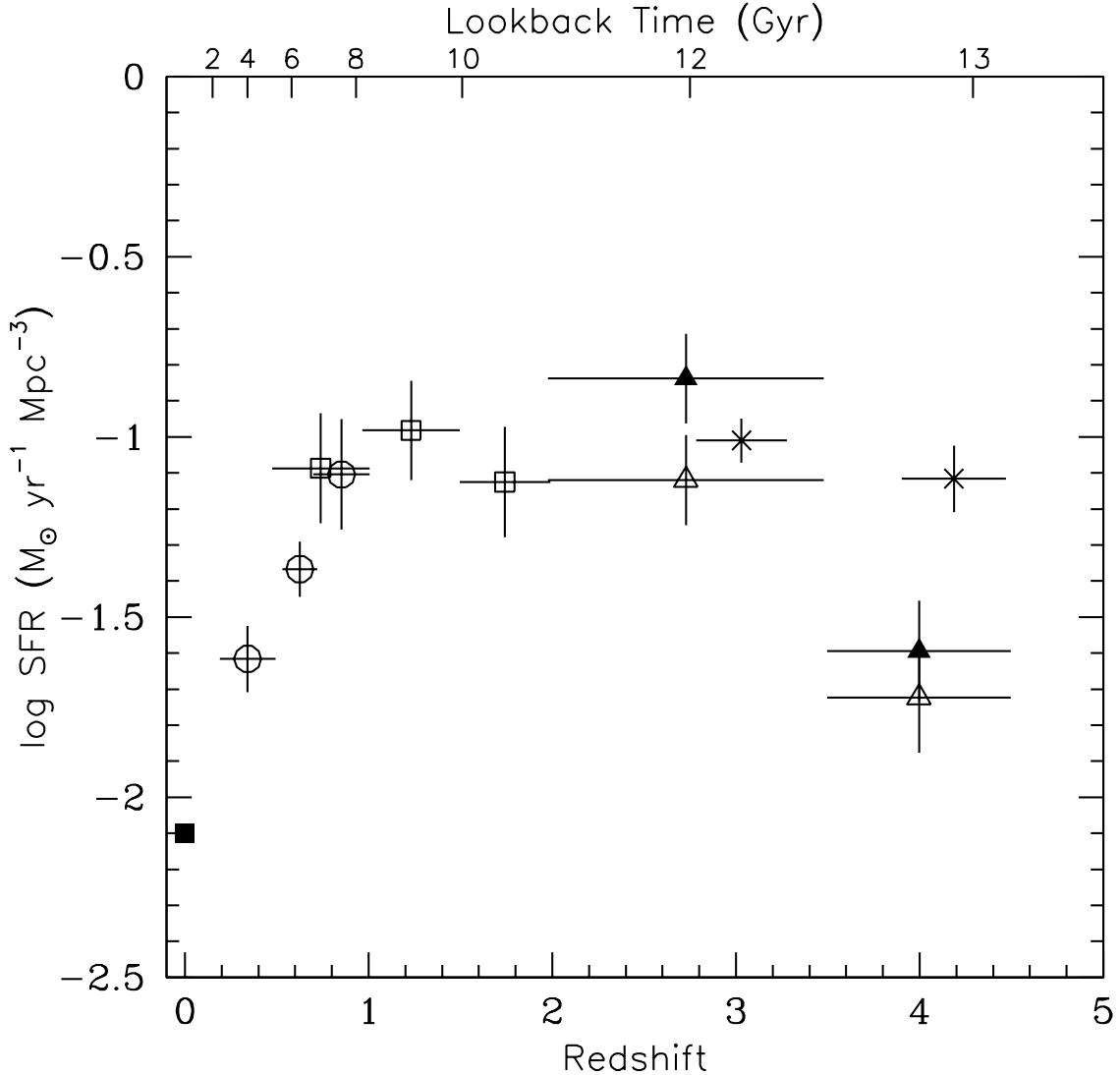}}
\fi
\caption{ \label{fig_lumden} \small 
({\it Left}) Star-formation rate
density vs. redshift derived from ultraviolet luminosity density. The $z > 2$
points are from Lyman-break objects in the Hubble deep field (HDF) north
({\it open triangles}), in the HDF-south
({\it filled triangles}) and in the Steidel et al. (1999) ground-based survey
({\it X}'s). The luminosity density has been determined by
integrating over the luminosity function and
correcting for extinction following the prescription of
Steidel et al. (1999).
Distances and volumes are computed using the cosmological parameters
$h,\Omega_{\rm m}, \Omega_\Lambda, \Omega_{\rm tot} = 0.65, 0.3, 0.7, 1.0.$
Possible contributions from far-IR and
sub-millimeter sources are not included. Also not included are the upward
revisions of the $z < 1$ star-formation densities suggested
by Tresse \& Maddox (1998) and Cowie et al. (1999). For the
lower-redshift points, the {\it open squares} are from HDF photometric
redshifts by Connolly et al (1997), the {\it open circles} are
from Lilly et al. (1996), and the {\it solid square} is from the H$\alpha$
survey of Gallego et al. (1995).
}
\end{figure}

\nocite{GZAR95}

\fi

Photometric redshifts computed via template fitting can also be used
to identify samples of high-$z$ objects, and
several such samples have been presented and discussed 
\citep{LYF96,SLY97,SY98,MP98,Rowan-Robinson99,FontanaHDFS}. 
Luminosity densities
at $z > 2$ from these studies are generally higher than those of \citet{MFDGSF96},
and the apparent drop in $\MER$ at $z \gtrsim 2$ is not universally
evident. A striking example of the difference between color-selected
samples and photometric-redshift selected samples is the fact
that HDF-S comes out with a higher UV luminosity-density 
than HDF-N in the color-selected sample shown in Fig. \ref{fig_lumden},
but a lower luminosity density in the photometric-redshift
selected sample of \citet{FontanaHDFS}.
One general concern about object selection is that galaxies
at lower redshift are typically {\it brighter} than the high-$z$
objects. Interlopers that make it into photometric samples 
can dominate the UV luminosity-density estimates. 

It does appear evident that the original criteria of \citet{MFDGSF96} were too 
conservative, and the {\it directly measured} UV luminosity density 
in bins at $z \sim 3$ and $z \sim 4$ is not demonstrably lower
than it is at $z \sim 1$. 
At still higher redshifts, from the very small number
of candidate objects with $z > 5$, \citet{LYF98}
computes an upper limit on the star-formation density at $z \sim 5 - 12$.
For the cosmology adopted here, the limit on star formation in
galaxies with $\dot{M} \gtrsim 100 h \msol {\rm yr^{-2}}$ is
lower than the inferred rate from Lyman-break galaxies at $z = 4$. 
With NICMOS observations it has become possible to identify
$z > 5$ candidates to much fainter limits, detecting objects
with UV luminosities more like those of typical Lyman break
galaxies at $2 < z < 5$.  \citet{Dickinson00} derives number
counts for color--selected candidates at $4.5 < z < 8.5$ in the
HDF--N/NICMOS survey, and finds that they fall well below
non--evolving predictions based on the well--characterized $z=3$
Lyman break luminosity function.  It appears that the space
density, the luminosities, or the surface brightnesses (and hence,
the detectability) of UV bright galaxies fall off at $z > 5$,
at least in the HDF--N.

\subsubsection{Dust and selection effects} \label{sec_dust}

Many of the discussions of the Madau diagram have centered around
selection effects and whether the data actually support a decrease in
$\MER$ for $z \gtrsim 2$. Star-formation in dusty or 
low-surface-brightness galaxies may be unaccounted
for in the HDF source counts.  Meurer and collaborators \citep{MHLLL97,MHC99} use local starburst galaxy samples to calibrate a
relation between UV spectral slope and far-IR ($60-100\mu$m) 
emission, and hence compute bolometric corrections for dust
attenuation. They apply these corrections to a sample of
color-selected Lyman-break galaxies in the HDF and derive an
absorption-corrected luminosity density at $z \sim 3$ 
that is a factor of 9 higher than
that derived by \citet{MFDGSF96}. Not all of this comes from the dust
correction; the luminosity-weighted mean dust-absorption factor for the
\citet{MHC99} sample is $5.4$ at 1600{\AA}. \citet{SY98} analyzed
the optical/near-IR spectral energy distributions of spectroscopically
confirmed Lyman-break galaxies in the HDF and found the best fit
synthetic spectra involved corrections of more than a factor of 10.
Smaller corrections were derived by \citet{SAGDP99} for a sample
of galaxies of similar luminosity.
In general, the danger of interpreting UV spectral slopes as a
measure of extinction is that the inferred luminosity corrections are
very sensitive to the form of the reddening law at UV
wavelengths. For the \citet{CKS94} effective attenuation law, used in
all of the aforementioned studies, a small change in the UV color
requires a large change in the total extinction.  Photometric errors
also tend to increase the dust correction, because objects that scatter
to the red are assigned larger corrections that are not offset by
smaller corrections for the blue objects.  The corrections for dust
must thus be regarded as tentative, and must be confirmed with more
extensive H$\alpha$ studies such as that of \citet{PKSDAG98}, and further studies in the mid-IR, radio, and sub-millimeter.


\ifsubmode\else
\begin{figure}
\ifsubmode\else
\centerline{\plotfiddle{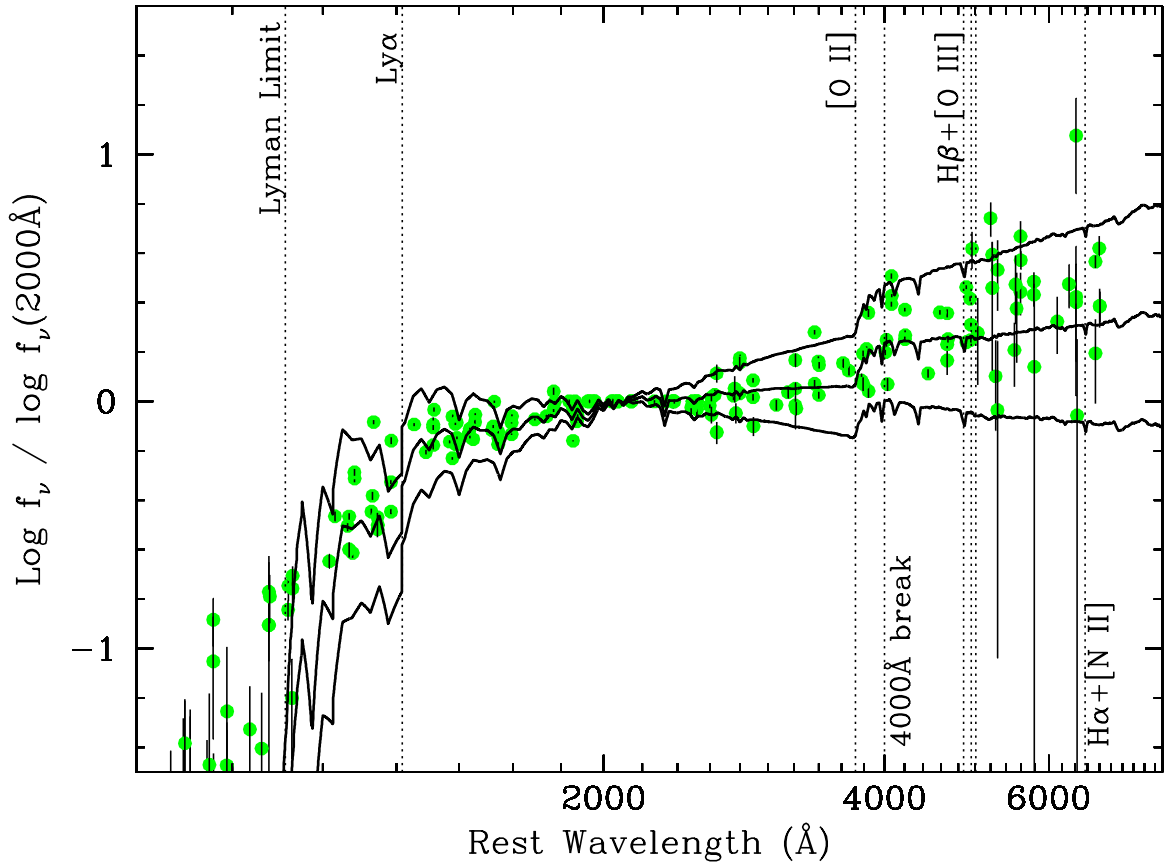}{3.0in}{-90}{60}{60}{-470}{380}}
\fi
\caption{ \label{fig_seds} 
Spectral energy distributions
of 28 spectroscopically confirmed Lyman-break galaxies with $2.5 < z < 3.5$,
superimposed on stellar population models. The models assume solar metallicity
and a Salpeter IMF and include attenuation due to the IGM for $z = 3$.
The models reflect constant star formation for $10^9$,
with different amounts of interstellar dust attenuation.
The reddening values are $E(B-V) = 0, 0.2,$ and $0.4$, with
the {\it curve} that is highest at long wavelengths having the highest
value. The Calzetti et al. (1994) attenuation law was used.
}
\end{figure}
\fi

The NICMOS HDF observations provide some additional insight into 
possible reddening corrections. Figure \ref{fig_seds} shows the 
spectral-energy distributions of HDF galaxies with spectroscopic
$2.5 < z < 3.5$, superimposed on models with a
constant star-formation rate and an age of $10^9$ years.
The galaxies are typically best fitted with ages $10^{7-9}$ years
and reddening $0.1 < E(B-V) < 0.4$ with the \citet{CKS94} attenuation law.
Very young galaxies with little extinction appear to be rare, and 
older or more reddened galaxies would possibly escape the Lyman-break
selection. However, \citet{Ferguson_photz} identified only 12 galaxies with
$\V < 27$ in the HDF-N that fail the \citet{Dickinson98p219} color criteria
but nevertheless have $2.5 < \zphot < 3.5$ from spectral-template fitting.
In general these galaxies fall just outside the color
selection boundaries, and the non-dropout sample contributes only 12\% of
the luminosity-density of the dropout sample.
It thus appears unlikely that there is a population that {\it
dominates} the metal-enrichment rate at $z \sim 3$ that is just
missing from the Lyman-break samples. If the optically unidentified 
sub-millimeter sources (see \S\ref{sec_submm1}) are really at $z \gtrsim 3$, 
they are probably a disjoint population, rather than simply the red tail of the 
Lyman-break population.

If sources akin to ultra-luminous IRAS galaxies were present 
at $z \sim 3$ in the HDF, it is likely that they would be detected,
but unlikely that they would be identified as Lyman-break galaxies.
\citet{TKS99} use HST observations to extend the 
spectral energy distributions for three nearby ultra-luminous
IR galaxies down to rest-frame 1400 {\AA}. Although
this is not far enough into the UV to predict HDF $B_{450}$ colors
at $z=3$ with confidence, it at least gives some indication
of whether such galaxies are detectable. All three
galaxies would likely be detected in the $F814W$ band if shifted out
to $z=3$, although VII Zw 031 would be very near the detection limit.
None of the galaxies would meet the \citet{MFDGSF96} Lyman-break
galaxy selection criteria, but might plausibly show up as high-$z$
objects from their photometric redshifts.

In addition to dust, cosmological surface-brightness can bias the
samples of high-redshift galaxies.
For fixed luminosity and physical size, i.e. no evolution,
surface-brightness will drop by about 1 mag between $z=3$ and
$z=4$. This lower surface brightness will result in a decrease in the
number density of objects and the inferred luminosity density, even if
there is no intrinsic evolution. Using simulated images for one
particular galaxy evolution model \citep{FB98}, \citet{fergHDFsymp}
arrived at corrections to $\MER$ of a factor of 1.5 at $z \sim 3$ and
4.7 at $z \sim 4$. 
\citet{Lanzetta_photz} have looked at
surface-brightness selection in a less model-dependent way, computing
the star-formation ``intensity'' (in 
solar masses per year per square kiloparsec)
from the UV flux in each pixel of the galaxy images.  The comoving volume
density of the regions of highest star-formation intensity appears to
increase monotonically with increasing redshift, whereas a strong
selection cutoff for star-formation intensities less than $1.5 \,\msol
\rm yr^{-1} kpc^{-2}$ affects samples beyond $z >2$. The results
suggest that surface-brightness effects produce a substantial
underestimate of $\MER$ at high redshift. 

The undetected low-surface-brightness UV-emitting regions should
contribute to the total diffuse background in the image. \citet{Bernstein97}
has attempted to measure the mean level of the extragalactic background
light (EBL) in the HDF images, while \citet{Vogeley97} has analyzed
the autocorrelation function of the residual fluctuations after masking
the galaxies. \citet{Vogeley97} concludes that diffuse light
clustered similarly to faint galaxies can contribute no more than 
$\sim 20$ \% of the mean EBL. In contrast \citet{Bernstein97},
concludes that the total optical EBL is two to three times the
integrated flux in published galaxy counts. The two results can
be made marginally compatible if the fluxes of detected
galaxies are corrected for light lost outside the photometric apertures
and for oversubtraction of background due to the overlapping wings of
galaxy profiles.  If this interpretation is correct,
there is not much room for the UV luminosity density to 
increase significantly at high redshift.

\subsubsection{Connection to galaxy evolution}

An important use of the metal-enrichment rate derived from the HDF and
other surveys is the attempt to ``close the loop'': to show that the
emission history of the universe produces the metal abundances and
stellar population colors we see at $z \sim 0$.  \citet{MFDGSF96} made
a first attempt at this, concluding that the metals we observe being
formed [integrating $\MER$ over time] are a substantial fraction of the
entire metal content of galaxies.  

\citet{MPD98} compared the integrated color of galaxy populations
in the local universe to that expected from the UV-emission history,
exploring a variety of options for IMF and dust obscuration.  With a
Salpeter IMF and modest amounts of dust attenuation, they find that a
star-formation history that rises by about an order of magnitude from
$z = 0$ to a peak at $z \sim 1.5$ is compatible with the present-day
colors of galaxies, with the FIR background, and with the metallicities
of damped Ly$\alpha$ absorbers. \citet{FCP96}, \citet{CH99} and
\citet{PFH99} have attempted to incorporate dust and
chemical evolution in a more self-consistent way. The models include
a substantial amount of obscured star-formation; more than 50\% of the
UV radiation is reprocessed by dust. 
The obscuration corrections increase the value of $\MER$ from samples 
already detected in optical surveys, but do not introduce whole
classes of completely dust-obscured objects.  Although there are significant
differences in the inputs and assumptions of the models, in all
cases a model with a peak in $\MER$ at $z \sim 1-1.5$ is found
consistent with a wide variety of observations. In particular the
models can simultaneously fit the COBE DMR and FIRAS measurements of the cosmic 
IR background \citep{PABBBDH96,Hauser_etal98,FDMBS98} 
and the integrated light from galaxy counts. The total mass in metals
at $z=0$ is higher in these new models than in those of \citet{MFDGSF96},
but the local census now includes metals in cluster X-ray gas, which
were ignored by \citet{MFDGSF96}.

Overall, the success of these consistency checks is 
quite remarkable. Various imagined populations of galaxies
(dwarfs, low-surface-brightness galaxies, highly dust obscured objects, etc.)
now seem unlikely to be cosmologically dominant. The fact that
the UV emission, gas metallicities, and IR backgrounds all
appear capable of producing a universe like the one we see today
leaves little room for huge repositories of gas and stars
missing from either our census at $z = 0$ or our census at 
high redshift.

Nonetheless, there is room for caution in this conclusion.  In clusters
of galaxies, the mass of metals ejected from galaxies into the X-ray
emitting gas exceeds that locked inside stars by a factor of 2-5
\citep{ML97}.  If the same factor applies to galaxies
outside clusters \citep{Renzini97}, then the local mass-density of metals
greatly exceeds the integral of metal-enrichment rate, implying that
{\it most} star-formation is hidden from the UV census (although the
differences in galaxy morphology in clusters and the field suggests
that clusters might not be typical regions of the universe).  Various
lines of evidence cited in \S\ref{sec_ellipticals} point to old ages
for elliptical galaxies (both inside and outside of clusters) and the
bulges of luminous early-type spirals \citep{Renzini99p9,GGJ99}.
The requirements for the early formation of metals in these systems
look to be at odds with the inferences from the models described above.
\citet{Renzini99p9} estimates that 30\% of the current density of metals
must be formed by $z \sim 3$, whereas the best-fit models to the
evolution of the luminosity density have only 10\% formed by then. The
discrepancy is interesting but is not outside of the range of error of
the estimates of both $\MER$ and the ages of stellar populations in
present-day spheroidal systems.

It also remains a challenge to ascertain how the metals
got from where they are at high redshift to where they are today. The
bulk of the metals locked up in stars at $z = 0$ are in luminous,
normal, elliptical and spiral galaxies. If elliptical galaxies (and
spheroids in general) formed early and rapidly, they probably account
for the lion's share of $\MER$ above $z=2$.  Thus the metals formed
in the $z = 1$ peak of $\MER$ must end up for the most part in
luminous spiral galaxies today. This is difficult to reconcile with
the lack of evolution observed in number-density or luminosity
of luminous spirals out to $z = 1$. An order-of-magnitude decline 
in star-formation rate in galaxy disks since $z = 1$ also seems
inconsistent with present-day colors of spiral galaxy disks, or with
star-formation histories derived from chemical-evolution models (e.g.
\citealt{Tosi96p299,PS98}). Furthermore, at $z \sim 1$ it
appears that compact-narrow-emission-line galaxies
\cite{LickHDFcompact2} and irregular galaxies account for a significant
fraction of the UV luminosity density. If these galaxies fade
into obscurity today, then they have not been accounted for in the
census of metals in the local universe, and the $z = 0$ metallicity
should be revised upwards in the global chemical-evolution models.
On the other hand if these galaxies are merging into luminous spirals
and ellipticals, it is hard to understand how luminous spiral and elliptical 
galaxy properties can remain consistent with PLE models out $z = 1.$

\subsection{Clustering of high-redshift galaxies}\label{sec_clustering}

Until relatively recently, evolution of the spatial two-point
correlation function $\xi$ has typically
been modeled as a simple power-law evolution in redshift 
\begin{equation}
\xi(r,z) = (r_0/r)^\gamma (1+z)^{-(3+\epsilon-\gamma)},
\end{equation}
where $r$ and $r_0$ are expressed in comoving coordinates
(e.g. \citealt{GP77}). For $\epsilon = 0$,
the formula above corresponds to stable clustering (fixed in 
proper coordinates), while for $\epsilon = \gamma - 3$ it
corresponds to a clustering pattern that simply expands with
the background cosmology. Although these two cases may bound the
problem at relatively low redshift, the situation becomes 
much more complex as galaxies surveys probe to high redshifts. 
In particular, the complex merging and fading histories of
galaxies make it unlikely that such a simple formula could
hold, and the differences in sample selection at high and
low-$z$ make it unclear whether the analysis is comparing
the same physical entities.
In practice it is likely that galaxies are biased
tracers of the underlying dark-matter distribution, with a bias
factor $b$ that is non-linear, scale-dependent,
type-dependent, and stochastic \citep{MBML99,DL99}. 

In hierarchical models, the correlation function $\xi(r)$ of halos on
linear scales is given by the statistics of peaks in a Gaussian random
field \citep{BBKS86}. Peaks at a higher density threshold $\delta \rho
/ \rho$ have a higher correlation amplitude, and their bias 
factor $b$, relative to the overall mass distribution is 
completely specified by the number-density of peaks (e.g \citealt{MM98}).

The comparison of the clustering of high-redshift galaxies in the HDF
and in ground-based surveys provides a strong test of whether there is
a one-to-one correspondence between galaxies and peaks in the
underlying density field. The bias factor observed in Lyman-break
samples \citep{SADGPK98,ASGDPK98,GSADPK98} is in remarkable agreement
with the expectations from such a one-to-one correspondence.  As the
luminosity of Lyman-break galaxies decreases, the number density
increases, and the clustering amplitude decreases \citep{LBG_clustering00}.
These trends are
all in agreement with a scenario in which lower-luminosity Lyman-break
galaxies inhabit lower-mass halos.  Thus, at redshifts $z \sim 3$, the
connection between galaxies and dark matter halos may in fact be quite
simple and in good agreement with theoretical expectations.

The picture becomes less clear for the samples described in
\S\ref{sec_clustering_phenomenology}. In particular, the interpretation
of the apparent evolution in the correlation length $r_0$ or the bias
factor $b$ from the \citet{MagMad99} and \citet{ACMMLFG99} studies
involves a sophisticated treatment of the merging of galaxies and
galaxy halos over cosmic time.  Fairly detailed attempts at this have
been made using either semi-analytic models \citep{BBCFL99,KCDW99} or more generic arguments (e.g. \citealt{MCLM97,MCLM98,MBML99}, following on earlier work by
\citealt{MF96}, \citealt{MW96}, \citealt{Jain97}, \citealt{ORY97} and others).  These studies all assume that
galaxy mass is directly related to halo mass, which may not be true in
reality at lower $z$, but which is an assumption worth testing.  A generic
prediction of the models is that the effective bias factor (the value
of $b$ averaged over the mass function) should increase with increasing
redshift. The correlation length $r_0$ is relatively independent of
redshift (to within factor of $\sim 3$), in contrast to the factor of
$\sim 10$ decline from $z=0$ to $z=4$ predicted for non-evolving bias.
The results shown by \citet{ACMMLFG99} are qualitatively consistent
with this behavior. However, it
is worth re-emphasizing that the HDF is a very small field.
The standard definition for the bias factor $b$ is the ratio of 
the root mean square density fluctuations of galaxies relative to mass
on a scale of $8 h^{-1} \Mpc$. Fluctuations on this angular scale 
clearly have {\it not} been measured in the HDF, and the interpretation
rests on ({\it a}) the assumption of a powerlaw index $\gamma = 1.8$ for
the angular correlation function (which is not in fact predicted
by the models) \citep{MCLM98}, and ({\it b}) the fit
to the integral constraint. Clearly much larger areas are needed
before secure results can be obtained.

\citet{RV97} and \citet{RVMB99} point out that much
of the measured clustering signal in the HDF comes from scales 
$25-250 \kpc$, which is within the size of a typical $L^*$ galaxy
halo at $z = 0$. The connection of the observed correlation function
to hierarchical models thus depends quite strongly on 
what happens when multiple galaxies inhabit the same halo.
Do they co-exist for a long time (e.g. as in present-day galaxy
groups and clusters), or rapidly merge together to form a larger
galaxy (in which case one should consider a ``halo exclusion 
radius'' in modeling the correlation function)? As the 
halo exclusion radius increases, the predicted correlation
amplitude on scales of $5\arcsec$ decreases relative to the standard
cosmological predictions. The slope of $\omega(\theta)$ also
differs from $\gamma = 1.8$ for $\theta \lesssim 20\arcsec$. 
\citet{RVMB99} find that the HDF data for $1.5 < \zphot < 2.5$
are best fit with stable clustering, a halo-exclusion radius
of $r_{\rm halo} = 200 h^{-1} \kpc$, and a low-density universe.

\section{Conclusion}
					
The HDFs represent an important portion of the frontier
in studies of the distant universe. Although we can point to few
problems that were solved by the HDF alone, the HDF has contributed
in a wide variety of ways to our current understanding of
distant galaxies and to shaping the debate over issues such as the
origin of elliptical galaxies and the importance of obscured star
formation. 
The scientific impact of the HDF can be attributed in part to the wide
and nearly immediate access to the data. The fact that many groups and
observatories followed this precedent (both for HDF data and
subsequently for data from other surveys) illustrates 
that the deeper understanding of the universe will come not from
any one set of observations but from sharing and comparing different
sets of observations.
Not all kinds of surveys are amenable to this kind of 
shared effort, but the precedent set by the HDF in the social aspects
of carrying out astronomical research may ultimately rival its significance
in other areas.

\section{Acknowledgments}
We are indebted to our fellow HDF enthusiasts, too numerous to mention, for
the many fruitful discussions that have provided input for this
review. The observations themselves would not have happened without
the dedicated contribution from the HST planning operations staff,
to whom we owe the largest debt of gratitude.
This work was based in part on obsevations with the NASA/ESA Hubble
Space Telescope obtained at the Space Telescope Science Institute,
which is operated by the Association of Universities for Research
in Astronomy, Inc., under NASA contract NAS5-26555.
This work was partially supported by NASA grants GO-07817.01-96A
and AR-08368.01-97A.

\bibliography{annrevmnemonic,bib} 
\bibliographystyle{annrev}


\ifsubmode
\begin{deluxetable}{rl}
\tablewidth{0pt}
\tablecaption{Census of objects in the central HDF-N}
\tablehead{
\colhead{Number}
 & \colhead{Type of source}
}
\startdata

$\sim 3000$ & Galaxies at $\U,\B,\V,\I$. \nl
$\sim 1700$ & Galaxies at $J_{110}, H_{160}$. \nl
$\sim 300$ & Galaxies at $K$ \nl
9 & Galaxies at $3.2\mu$m \nl
$\sim 50$ & Galaxies at 6.7 or 15$\mu$ \nl
$\sim 5$ & Sources at $850\mu$m \nl
$0$ & Sources at 450$\mu$m or 2800$\mu$m \nl
6 & X-ray sources \nl
$\sim 16$ & Sources at 8.5 GHz \nl
$\sim 150$ & Measured redshifts \nl
$\sim 30$ & Galaxies with spectroscopic $z > 2$ \nl
$<20$ & Main-sequence stars to $I = 26.3$ \nl
$\sim 2$ & Supernovae \nl
$0-1$ & Strong gravitational lenses \nl
\enddata
\end{deluxetable}
\fi

\clearpage
 

\ifsubmode

\begin{figure}
\caption{\label{fig_mendez}
Color-magnitude diagrams for point sources in the 
Hubble deep fields (HDF).
({\it Dash-dot line}) Locus of disk M-dwarfs at a distance of 8 kpc; 
({\it solid star symbols}) main-sequence disk and halo stars. ({\it Horizontal line})
The ($15\sigma$) detection limit of $I=27$ magnitude; ({\it filled circles})
galactic stars and unresolved distant galaxies.
({\it Open star symbols}) Point sources which are too faint to be on the
main sequence at any reasonable distance, and which may be white dwarfs (WD).
({\it Dashed line}) The
$0.6 \msol$ WD cooling track from Hansen (1999) for a distance of 2 kpc.
(Adapted from Mendez \& Minniti 2000)
}
\end{figure}

\begin{figure}
\caption{\label{fig_volume}
An illustration of volume and time in the Hubble deep fields (HDF). 
({\it Top})
For the 5~arcmin$^2$ WFPC2 field of view, the co-moving volume out
to redshift $z$ is plotted for several cosmologies, scaled
by the present day normalization of the galaxy luminosity function
[$\phi^\ast$, here taken to be 0.0166$h^3$~Mpc$^{-3}$ from
Gardner (1997)].  This gives a rough measure of the
number of ``$L^\ast$-volumes'' out to that redshift.  ({\it Bottom})
The fractional age of the universe versus redshift is shown.
Most of cosmic time passes at low redshifts, where the HDF volume
is very small.}
\end{figure}

\nocite{GSFC97}

\begin{figure}
\caption{\label{fig_morphology}
Selected galaxies from the Hubble deep field north, viewed at optical and near-infrared
wavelengths with WFPC2 and NICMOS.  For each galaxy, the 
color composites are made from $B_{450}$, $V_{606}$, $I_{814}$ ({\it left panel})
and $I_{814}$, $J_{110}$, $H_{160}$ ({\it right panel}).  The NICMOS images have 
poorer angular resolution. In the giant spiral at $z = 1.01$, a prominent {\it red bulge}
and {\it bar} appear in the NICMOS images, whereas the spiral arms and HII regions
are enhanced in the WFPC2 data.  For most of the other objects, including
the more irregular galaxies at $z \sim 1$ and the Lyman Break galaxies
at $z > 2$, the morphologies are similar in WFPC2 and NICMOS.
}
\end{figure}

\begin{figure}
\caption{\label{fig_counts}
Galaxy counts from the Hubble deep fields (HDF) and other surveys.
The HDF galaxy counts ({\it solid symbols}) use isophotal magnitudes and
have not been corrected for incompleteness. These corrections
will tend to steepen the counts at the faint end, but in a model-dependent
way. For the $K$-band, a color correction of $-0.4$ mag has been
applied to the NICMOS F160W band magnitudes. The HDF-N counts from
Thomson et al (1999) ({\it filled squares}) and the HDF-S
counts from Fruchter et al. 2000 ({\it filled circles}) are filled circles
are shown. For the
$U,B$ and $I$ bands, the HDF counts are the average of HDF-N
(Williams et al. 1996) and HDF-S (Casertano et al. 2000) with
no color corrections. The groundbased counts ({\it open symbols})
are from Mcleod \& Rieke (1995), Gardner et al. (1993; 1996),
Postman et al. (1998), Lilly Cowie \& Gardner (1991),
Huang et al. (1997), Minezaki et al. (1998), Bershady (1998), Moustakas et al. (1997),
and Djorgovski et al. (1995). ({\it Smooth curves}) A
no-evolution model with $\Omega_M,\Omega_\Lambda,\Omega_{\rm tot} =
0.3, 0.7, 0.1$ based on the luminosity functions, spectral-energy
distributions, and morphological type mix of the ``standard NE''
model of Ferguson \& McGaugh (1995).
}
\end{figure}
 
\nocite{MR95,LCG91,GCW93,GSCF96,HCGHSW97,MKYP98,BLK98,MDGSPY97,DSPLSNSMHBCHN95,FM95}
 
\begin{figure}
\caption{ \label{fig_lumden} \small ({\it Left}) Star-formation rate
density vs. redshift derived from ultraviolet luminosity density. The $z > 2$
points are from Lyman-break objects in the Hubble deep field (HDF) north 
({\it open triangles}), in the HDF-south
({\it filled triangles}) and in the Steidel et al. (1999) ground-based survey
({\it X}'s). The luminosity density has been determined by
integrating over the luminosity function and
correcting for extinction following the prescription of
Steidel et al. (1999). 
Distances and volumes are computed using the cosmological parameters
$h,\Omega_{\rm m}, \Omega_\Lambda, \Omega_{\rm tot} = 0.65, 0.3, 0.7, 1.0.$
Possible contributions from far-IR and
sub-millimeter sources are not included. Also not included are the upward
revisions of the $z < 1$ star-formation densities suggested
by Tresse \& Maddox (1998) and Cowie et al. (1999). For the
lower-redshift points, the {\it open squares} are from HDF photometric
redshifts by Connolly et al (1997), the {\it open circles} are
from Lilly et al. (1996), and the {\it solid square} is from the H$\alpha$
survey of Gallego et al. (1995).
}
\end{figure}
 
\nocite{GZAR95}

\begin{figure}
\caption{ \label{fig_seds} Spectral energy distributions
of 28 spectroscopically confirmed Lyman-break galaxies with $2.5 < z < 3.5$,
superimposed on stellar population models. The models assume solar metallicity
and a Salpeter IMF and include attenuation due to the IGM for $z = 3$.
The models reflect constant star formation for $10^9$,
with different amounts of interstellar dust attenuation.
The reddening values are $E(B-V) = 0, 0.2,$ and $0.4$, with
the {\it curve} that is highest at long wavelengths having the highest
value. The Calzetti et al. (1994) attenuation law was used.
}
\end{figure}
\fi

\end{document}